\documentclass[fleqn,usenatbib]{mnras}

\usepackage{newtxtext,newtxmath}
\usepackage[T1]{fontenc}
\usepackage{ae,aecompl}

\usepackage{epsf}
\usepackage{cleveref}

\crefformat{section}{\S#2#1#3} % see manual of cleveref, section 8.2.1
\crefformat{subsection}{\S#2#1#3}
\crefformat{subsubsection}{\S#2#1#3}

\usepackage{graphicx}	% Including figure files
\usepackage{epstopdf}
\usepackage{amsmath}	% Advanced maths commands
\usepackage{siunitx}
\usepackage{longtable}
\usepackage{booktabs}
\usepackage{multirow}
\usepackage{multicol}
\usepackage{makecell}
\usepackage{supertabular}
\usepackage{sansmath}

\newcommand{\Halpha}{H{$\alpha$}}

\newcommand{\OIII}{[O\,{\sc iii}]}

\newcommand{\taunu}{\tau_{\nu}}

\newcommand{\nrmse}{\mathsf{nrmse}}
\newcommand{\spshell}{{\it sp shell}}
\newcommand{\plshell}{{\it pl shell}}
\newcommand{\thetaERD}{$\theta_{\mathrm{erd}}$}

\newcommand{\nucr}[1]{$\nu_{cr}$}

\newcommand{\NoFittedVLAImages}{123} %total images accessed for diams
\newcommand{\NoNewVLADiams}{82} %number of newly fitted angular diams in VLA
\newcommand{\NoVLAOverlap}{49} %number of fitted angular diams already in the literature
\newcommand{\NoNewHSTDiams}{120} %number of newly fitted angular diams in HST
\newcommand{\NoWithRadio}{1141} %number of pne with at least 1 flux density in different bands
\newcommand{\NoFoundObjects}{391} %number of pne with at least 3 flux densities in different bands
\newcommand{\NoSEDfittedspshell}{262} %number of pne with succesful SED fit for sp shell
\newcommand{\NoSEDfittedplshell}{237} %number of pne with succesful SED fit for pl shell
\newcommand{\noRLandspshelldiams}{130} %number of pne with both spshell and RL diam
\newcommand{\noRLandplshelldiams}{144} %number of pne with both plshell and RL diam
\newcommand{\noOLandspshelldiams}{152} %number of pne with both spshell and OL diam
\newcommand{\noOLandplshelldiams}{149} %number of pne with both plshell and OL diam
\newcommand{\RadioReferences}{Lynds1963,
Menon1965,
Slee1965,
Terzian1966,
Thomasson1970,
Aller1972a,
Terzian1973,
Milne1975,
Milne1979,
Kwok1981,
Mross1981,
Calabretta1982,
Milne1982,
Purton1982,
Gathier1983,
Seaquist1983,
Isaacman1984,
Reich1984,
Kwok1985a,
Rodriguez1985,
Basart1987,
Phillips1988,
Zijlstra1989a,
Aaquist1990,
Furst1990,
Ratag1990,
Aaquist1991,
Becker1991,
Ratag1991,
Kwok1993a,
Pottasch1994,
Wright1994,
Steene1995,
Douglas1996,
Taylor1996,
Rengelink1997,
Condon1998,
Condon1998a,
Garwood1998,
VandeSteene2001,
DeBreuck2002,
Mauch2003,
Nord2004,
Helfand2006,
Murphy2007,
Cerrigone2008,
Pazderska2009,
Urquhart2009,
Bojicic2010a,
Murphy2010a,
bojicic2011,
Hoare2012,
McConnell2012,
hurleywalker2017,
Intema2017,
Meyers2017,
Harvey-Smith2018,
hurleywalker2019} %radio references
\title[PNe angular diameters from SED modeling]{Determination of Planetary Nebulae angular diameters from radio continuum Spectral Energy Distribution modeling}

\author[I. S. Boji\v ci\'c et al.]{
I. S. Boji\v ci\'c$^{1}$\thanks{E-mail: ibojicic@gmail.com},
M. D. Filipovi\'c$^{1}$,
D. {Uro{\v s}evi{\'c}}$^{2,3}$,
Q. A. Parker$^{4,5}$ and
T. J. Galvin$^{6,7}$\\
$^1$Western Sydney University, Locked Bag 1797, Penrith South DC, NSW 1797, Australia\\
$^2$Department of Astronomy Faculty of Mathematics, University of Belgrade Studentski trg 16, 11000 Belgrade, Serbia\\
$^3$Isaac Newton Institute of Chile Yugoslavia Branch, Serbia\\
$^4$ The Laboratory for Space Research, Cyberport 4, University of Hong, Hong Kong\\
$^5$ The Department of Physics, CYM building, University of Hong Kong, Pok Fu Lam Road, Hong Kong
\\
$^6$ ICRAR ,International Centre for Radio Astronomy Research, Curtin University, Bentley, WA 6102, Australia\\
$^7$ CSIRO Astronomy and Space Science, PO Box 1130, Bentley WA 6102, Australia}

\date{Accepted XXX. Received YYY; in original form ZZZ}

\pubyear{2015}

\begin{document}
\label{firstpage}
\pagerange{\pageref{firstpage}--\pageref{lastpage}}
\maketitle

\begin{abstract}
Powerful new, high resolution, high sensitivity, multi-frequency, wide-field radio surveys such as the Australian Square Kilometre Array Pathfinder (ASKAP) Evolutionary Map of the Universe (EMU) are emerging. They will offer fresh opportunities to undertake new determinations of useful parameters for various kinds of extended astrophysical phenomena. Here, we consider specific application to angular size determinations of Planetary Nebulae (PNe) via a new radio continuum Spectral Energy Distribution (SED) fitting technique. We show that robust determinations of angular size can be obtained, comparable to the best optical and radio observations but with the potential for consistent application across the population. This includes unresolved and/or heavily obscured PNe that are extremely faint or even non-detectable in the optical.
\end{abstract}


%%%%%%%%%%%%%%%%%%%%%%%%%%%%%%%%%%%%%%%%%%%%%%%%%%

%%%%%%%%%%%%%%%%% BODY OF PAPER %%%%%%%%%%%%%%%%%%

\section{Introduction}

The estimation of accurate angular diameters is a crucial task in studies of Planetary Nebulae (PNe). Most of the essential physical parameters (e.g. physical size or distance, ionised mass and density) can be calculated only if the angular diameter is well determined. The angular diameter as seen in any waveband is an observational property and as such its accuracy strongly depends on the measurement method and the sensitivity of the chosen waveband to different physical characteristics of the object under study. For example, a narrow-band \OIII\ observation of a PN can give a very different value for angular diameter compared to a \Halpha\ equivalent due to the ionisation stratification that is present and that itself depends on the effective temperature of the ionising star, e.g. \cite{Kovacevic2011}.

Several methods have been proposed and used for representative determination of PN angular diameters. The most commonly used method is by measuring the maximum extent of the nebula at some prescribed brightness level in the considered wavelength range. This level is usually taken to be 10\% of the peak brightness \citep{Bensby2001} and can give good results for well resolved symmetric nebulae (in comparison with the resolution of the instrument used to obtain the measure). In these cases, there is a well defined contrast between the emitting and non-emitting regions and negligible self-absorption in the used observational band. It is usually applied to observations in the optical bands. The major advantage over other methods is the simplicity of application and non dependence on any assumption of the underlying structure of a PN. However, for objects whose projected angular size is approaching the point spread function (PSF) of the instrument from the telescope and atmosphere combination, the method will start to systematically overestimate true angular diameters. For example, this can be seen clearly when Hubble Space Telescope (HST) determined PNe angular sizes are compared to those from ground based measurements of compact PNe. The prescribed brightness level method above is currently not applicable for the majority of distant Galactic and almost all extragalactic PNe  as the projected angular sizes are too small compared to a typical ground based PSF at even the best sites, apart from those imaged by HST in the Magellanic Clouds \citep{Stanghellini2003, Shaw2006}.

The second commonly used method is the Gaussian deconvolution method. It is often used with radio observations in cases where the emission source is partially resolved and where an estimate can be obtained by comparison of the convolving Full Width at Half-Maximum power (FWHM) beam and its apparent FWHM. This method is applicable to measurements in optical bands (see for example \cite{Tylenda2003}). The basic method will work adequately only in the case of a Gaussian beam (PSF) and a Gaussian brightness distribution. In more complex circumstances, a conversion factor, related to observational parameters and intrinsic surface brightness distribution, must be adopted \citep{Bedding1994,Tylenda2003,Hoof2000a}. 

In the case of unresolved objects these previous methods obviously breakdown. In such cases only an upper limit for the angular diameter can be estimated. This situation heavily affects studies of compact/young and distant Galactic PNe and of course all extragalactic PNe. Having an accurate angular size estimate, especially for compact PNe, will provide a more reliable estimate of PN physical extent. This depends on more reliable distance determinations such as from the SB-r relation of \cite{Frew2016} (FP16 hereafter)  or from GAIA distances \citep{Kimeswenger2018} to their central stars (CSPN) when such CSPN are detectable (note many PNe have CSPN beyond GAIA limits). This then also helps with dynamical age estimates and evolution.

The second problem is related to the future angular diameter estimates of not-yet-detected Galactic PNe (GPNe). The current number of known GPNe is $\sim3.5\times10^3$ as compiled in the Hong Kong/AAO/Strasbourg H$\alpha$ (HASH) Planetary Nebula database and research platform \citep{Parker2016,Bojicic2017} while estimates of the total number of GPNe vary from 10\,000 \citep{Jacoby1980} to 46\,000 \citep{Moe2006}. Although the majority of known GPNe ($\sim90$\%; \citealt{Frew2010a}) are detected and successfully observed in optical (narrow) bands, this method is probably reaching its limits. This is despite considerable numbers still being found within and outside the Galactic Plane by a very active amateur astronomy community. They are fastidiously trawling extant on line surveys and even performing their own spectroscopic follow-up and ultra deep imaging \citep{Kronberger2016} and French amateurs (e.g. Le Du \& Acker, L'Astronomie No.68, January 2014 and Le Du, L'Astronomie No.133, March 2019). Our current understanding of the field predicts that the majority of not-yet-detected GPNe are either too distant, too evolved (and so of very low surface brightness) or lie at low galactic latitudes and are heavily obscured in optical bands by interstellar dust, or a combination of all factors (but see \cite{stenborg2017} based on deep, wide-field camera, CTIO 4m [OIII] imaging of the Galactic Bulge taken by author Parker).

New detection and confirmation methods have been employed in recent years \citep{Gledhill2018, Fragkou2018, Irabor2018, Anderson2012, Parker2012}. Most, if not all, of these methods are based on radio and mid-infrared properties of PNe. Radio continuum based measurements in particular are expected to become a critical tool in future studies as they are effectively insensitive to the presence of intervening dust. Hence, the next generation of high resolution interferometric radio continuum all-sky surveys will likely become the pre-eminent tool to estimate angular diameters of PNe. 

In this paper, we present a new method for determining angular sizes of PNe from modelled radio continuum spectral energy distributions (SEDs). The proposed method has an advantage over commonly used methods in its ability to accurately estimate the angular diameter of unresolved PNe. To showcase the accuracy and applicability of this method we used angular sizes previously reported for catalogued GPNe found in the literature or directly measured from available high resolution optical and radio images.

\section{Determination of angular diameters from radio continuum SED models}

To a first approximation, a PN can be considered as a static, ionised and homogeneous gas sphere with an exciting star placed in its centre. In this case we expect the spectral distribution of the emitted flux $S_{\nu}$ from a PN to have a relatively simple shape described with two distinct regions. In the limit of small opacities, the optically thin flux will become almost independent of the frequency ($S_{\nu}\propto\nu^{-0.1}$) while for cases of optically thick emission radio spectra will follow the black body emission distribution ($S_{\nu}\propto\nu^{2}$)

In this section we extend the consideration of spectral energy distribution in the radio regime from \cite{Olnon1975} and \cite{Pottasch1984}.

For PN models with axial symmetry the radio continuum flux density will have a frequency dependence of the form:

\begin{equation}
S_{\nu} = \frac{4\pi k T_e \nu^2}{c^2 D^2}\int_{0}^{\infty}\rho(1-e^{-\taunu(\rho)})d\rho
\label{eq:flux_nu}
\end{equation}
where $D$ is the distance, $\rho$ is the radius from the centre and $\tau(\rho)$ is the optical thickness at $\rho$. For assumption of a pure hydrogen, isothermal plasma the optical thickness can be approximated with:

\begin{equation}
\tau(\rho)=8.235 \times 10^{-2}\  T_e^{-1.35}\ \nu^{-2.1} \int n_e(r)^2 ds
\label{eq:tau}
\end{equation}
where electron temperature ($T_e$) and frequency ($\nu$) are in units of K and GHz, respectively and $n_e(r)$ is the electron density at distance $r$ from the centre of the nebula (in cm$^{-3}$). The integral $\textrm{EM}(\rho) = \int n_e(r)^2 ds$ is usually called the emission measure (EM) and it describes the effect of ionised particles along the line of sight ($s$).

If we know the critical frequency (\nucr\ ) at the centre of the nebula, i.e. the frequency where the radio emission changes from optically thin to optically thick ($\tau(0) = 1$), the emission measure through that point ($\textrm{EM}(0)$) can be calculated from Equation~\ref{eq:tau}.

\subsection{Spherical Shell}
 \label{sec:spshell}

In this paper we considered two simplified models of PNe density distributions. The first model will be the model of a constant density spherical shell defined with the outer radius (R) and the relative thickness of the shell ($\mu=$R$_{in}/$R) i.e. the ratio between the inner (R$_{in}$) and the outer radius. Following \cite{Olnon1975} we introduce $x=\rho/R$. Then the integral in the Equation~\ref{eq:flux_nu} can be written as:

\begin{equation}
\int_{0}^{\infty}\rho(1-e^{-\taunu(\rho)})d\rho = R^2\int_{0}^{\infty} x (1 - e^{-\taunu\cdot g_1(x)})dx
\label{eq:modelcylinder}
\end{equation}
where $\taunu$ is again the optical thickness trough the centre of the nebula and the geometry function $g_1(x)$ is defined as:

\begin{equation}
\begin{split}
g_1(x) 	& = \sqrt{1-x^2} - \sqrt{\mu^2-x^2}	& \text{for}~x < \mu, \\
		& = \sqrt{1-x^2}					& \text{for}~\mu\leq x < 1~\text{and} \\
		& = 0						&\text{for}~x \geq 1, \\
\end{split}
\end{equation}

We will refer to this model as \spshell.

It is important to note that the constant gas density stratification throughout the ionised part of the nebula will be too simplistic for some nebulae. Gas density stratification in a PN is initially a result from the strong ejection of gas layers with different masses, velocities and chemical compositions. It will continue to change during the subsequent evolution because of the energy-input from the fast wind and due to the progress of strong ionisation fronts throughout the neutral gas \citep{Perinotto2004}.

\subsection{Spherical Shell with a Power-Law Outflow}
 \label{sec:plshell}

Secondly, we also considered a truncated power-law distribution (extension of the Model V in \cite{Olnon1975}). If we consider the simplest case of uniform mass flow rate with constant velocity, then the gas density distribution, throughout the resulting layers, will behave as $\sim r^{-2}$. This will subsequently produce a $S_{\nu}\sim\nu^{0.6}$ spectrum in the optically thick part. We adopted this model with modification for a spherical shell instead of a filled sphere in the centre. With this modification the $g_2(x)$ function from OL75 will become:

\begin{equation}
\begin{split}
g_2(x)  & = 8/3 - 2\ \mu  \quad \text{for}~x = 0, \\
 	& = x^{-3}\Big[\frac{\pi}{2}-\arctan{\Big(\frac{1}{x}\sqrt{1-x^2}\Big)} - x\sqrt{1-x^2}\Big]+ \\
	& 2\Big(\sqrt{1-x^2}-\sqrt{\mu^2-x^2}\Big) \\
	& \quad \text{for}~0 < x\leq\mu,\\
	& =x^{-3}\Big[\frac{\pi}{2}-\arctan{\Big(\frac{1}{x}\sqrt{1-x^2}\Big)} - x\sqrt{1-x^2}\Big]+\\
	& 2\sqrt{1-x^2} \\
	& \quad \text{for}~\mu < x < 1~\text{and}\\
	& =\frac{\pi}{2}x^{-3}  \quad \text{for}~x \geq 1, \\
\end{split}
\end{equation}
and the flux density can be calculated from OL75 (Eq.~23). In the further text this model will be referred to as \plshell. In this case the estimate of the angular diameter will define the size of the spherical shell from the centre of the nebula. In principle, the full angular diameter from this method is bound only by the integration limits and the correction needs to be applied for meaningful comparison with other methods. From the $g_2(x)$ function we estimated that the emission in this model falls to 10\% of the peak value at 184\% of the outer diameter of the embedded spherical shell. Therefore, a correction of 1.84 was applied to the output fitting values. A similar correction for the \spshell\ model is not necessary. 

Here, we emphasise that the simplification of uniform mass flow rate will produce an extreme case of a PNe spectrum i.e. the optically thick SED region in the \plshell\ model will always have power law index of 0.6. From detailed hydrodynamical considerations of the coupled central star-gas system by numerous authors (e.g.~\cite{Volk1985,Schmidt-Voigt1987,Mellema1994,Marten1991,Marigo2001}) it is expected that the observed density distributions will highly depend on various factors like the precursor's mass-loss history, start and duration of the interaction with the fast wind from the central star and the position of the ionisation front. Thus, in the optically thick part of the spectrum we can expect all values of the power low index to be between --0.6 and --2.

\begin{figure}
\includegraphics[width=\columnwidth]{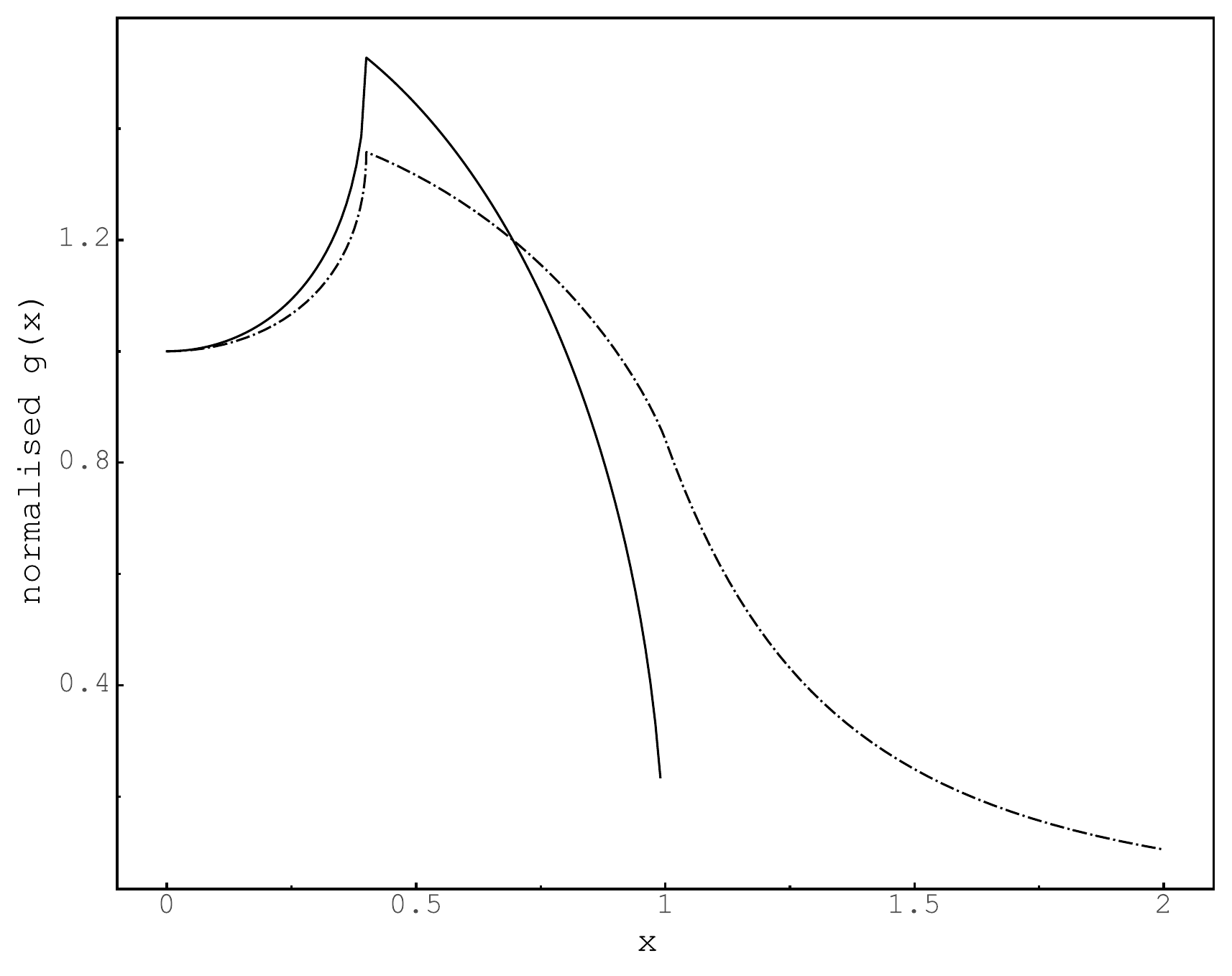}
\caption{Geometry functions (g(x)) used in the modelling for $\mu=0.4$. Solid and dashed line represent \spshell\ and \plshell\ model, respectively. Curves are normalised to have g(0) = 1.}
\label{fig:gxmodels}
\end{figure}

We present two geometry functions: $g_1(x)$ for the \spshell\ and $g_2(x)$ for the \plshell\ models in Figure~\ref{fig:gxmodels}.

As can be seen from Equations~\ref{eq:flux_nu}, \ref{eq:tau}, and \ref{eq:modelcylinder} the adopted models are defined with four free parameters: electron temperature ($T_e$), solid angle ($\Omega$), emission measure (EM) through the centre of the nebula and the relative thickness ($\mu$) for the \spshell\ and the \plshell\ model. We fixed electron temperature to its canonical value ($T_e=10^4$~K) and $\mu=0.4$ as this is the expected average value for majority of Galactic PNe \citep{Marigo2001,Schonberner2007}. The other two free parameters ($\Omega$ and EM) can be now estimated from nonlinear regression fitting using available radio data.

Since our method depends, observationally, largely on how much of the intrinsic flux is sampled, we dubbed the estimated diameter from the presented method as the {\it effective radio angular diameter} (\thetaERD).

\subsection{Fitting Procedure}

For the nonlinear fitting of SED models we used the {\it Trust Region Reflective} (TRF) algorithm implemented in the {\sc Python's} \textsf{ scipy.optimize.curve\_fit} module \citep{branch1999,virtanen2020}. For the $\Omega$ initial parameter we adopted the smallest non-resolving beam size multiplied by a factor randomly selected between numbers 2 and 4. i.e. we selected the smallest beam size still larger than the angular diameter found in the literature. The randomising factor was applied in order to avoid possible bias from observational methods (i.e. since a good radio-continuum interferometric observing procedure to measure the total flux from an object is to select a beam size just slightly higher than the expected source size in order to resolve the source from the nearby emission on one hand and avoid filtering out object's larger structures on the other). For the initial critical frequency ($\nu_{init}$) we used 1~GHz for all objects. We provide a {\sc Python} package \textsf{rtsed}\footnote{\href{https://github.com/ibojicic/rtsed}{github.com/ibojicic/rtsed}} with methods used in this paper to contribute to reproducible research efforts. 

In the first iteration we removed possible outliers using Grubb's test \citep{Grubbs1969}. The points removed with this method are mainly high resolution and high frequency observations ($\nu_{obs}>5$~GHz) commonly affected by undersampling. 

\begin{table*}
\centering
\caption{Newly measured angular diameters from NVAS images. The columns are as follows: (1) HASH PN ID; (2) PNG designation as defined in \citep{Acker1992a}; (3) radio band of the source image; (4), (5) and (6) estimated major axis, minor axis and positional angle, respectively; (7) and (8) beam parameters in the source image; (9) intensity level at 99.5 percentile, (10) intensity level at fitted angular diameter and (11) the date when the observation was made. {\bf Note: the positional angle was measured from the North Celestial Pole in the counterclockwise direction.}}
\label{tab:vlafitteddiams_short}
\begin{tabular}{rlrrrrcrrrl}
\toprule
(1) & (2) & (3) & (4) & (5) & (6) & (7) & (8) & (9) & (10) & (11) \\
  \#PN  & PNG & Band    & $\theta_{maj}$ & $\theta_{min}$ & PA$_{\theta}$ &                $\theta_{beam}$ & PA$_{beam}$ &   Peak &  10\% &     ObsDate \\
 &  &  & (") & (") & ($^{\circ}$) & ("$\times$") & ($^{\circ}$) & (mJy/beam)  & (mJy/beam) &             \\
\midrule
\multirow{2}{*}{23} & \multirow{2}{*}{000.3+12.2} & 3cm &            6.7 &            6.7 &            -6 &    0.92$\times$0.39 &         -65 &   3.14 &  0.34 &  2001-02-18 \\
    &            & 6cm &            5.4 &            5.0 &            36 &    0.63$\times$0.36 &          -5 &   1.88 &  0.22 &  1984-12-29 \\
\cline{1-11}
\cline{2-11}
68  & 001.5-06.7 & 3cm &            1.7 &            1.7 &            14 &    0.87$\times$0.22 &          29 &  30.26 &  3.28 &  1991-09-21 \\
94  & 002.4+05.8 & 6cm &           40.6 &           40.5 &           -50 &    7.94$\times$4.17 &          -3 & 115.41 & 12.97 &  1984-04-27 \\
\multirow{2}{*}{113} & \multirow{2}{*}{003.1+02.9} & 3cm &            8.1 &            7.9 &           -27 &    0.89$\times$0.42 &         -63 &   4.12 &  0.42 &  2001-02-18 \\
    &            & 6cm &            7.3 &            7.0 &           -30 &    0.72$\times$0.35 &         -12 &   3.77 &  0.36 &  1985-02-23 \\
\cline{1-11}
\cline{2-11}
123 & 003.5-04.6 & 3cm &            9.7 &            9.6 &            -4 &    0.67$\times$0.53 &          83 &   0.43 &  0.05 &  2001-02-18 \\
% 153 & 004.9+04.9 & 6cm &            3.2 &            3.1 &            29 &    0.67$\times$0.36 &         -12 &   3.17 &  0.27 &  1984-12-29 \\
% 186 & 006.7-02.2 & 6cm &           74.5 &           72.4 &            -2 &  26.77$\times$12.96 &          15 &  92.01 &  9.42 &  1984-08-05 \\
% 195 & 007.2+01.8 & 6cm &            5.7 &            5.7 &             5 &    0.63$\times$0.36 &          -2 &   5.17 &  0.66 &  1985-02-23 \\
... & ... & ... & ... & ... & ... & ... & ... & ... &  ... \\

\bottomrule
\end{tabular}
\end{table*}

\begin{table*}
\centering
\caption{Newly measured angular diameters from HST images. The columns are as follows: (1) HASH PN id ; (2) PNG designation as defined in \citep{Acker1992a}; (3), (4) and (5) estimated major axis, minor axis and positional angle, respectively; (6) intensity level at 99.5 percentile, (7) intensity level at fitted angular diameter, and (8) HST filter.}
\label{tab:hstfitteddiams_short}
\begin{tabular}{rlrrrrrc}

\toprule
(1) & (2) & (3) & (4) & (5) & (6) & (7) & (8) \\
\#PN & PNG & $\theta_{maj}$ & $\theta_{min}$ & PA$_{\theta}$ & Peak & 10\% &   Filter \\
 & & (") & (") & ($^{\circ}$) & (counts/s) & (counts/s)  &        \\

\midrule
23       &  000.3+12.2 &            7.3 &            7.1 &             5 & 18.30 &   1.48 &  f656n \\
74       &  001.7-04.4 &            2.4 &            2.4 &            76 &  5.34 &   0.43 &  f656n \\
90       &  002.2-09.4 &            6.6 &            6.6 &            -8 &  8.69 &   0.65 &  f656n \\
94       &  002.4+05.8 &           35.1 &           35.0 &           -65 &  4.14 &   0.40 &  f656n \\
105      &  002.7-04.8 &           10.8 &           10.7 &            51 &  2.13 &   0.18 &  f656n \\
% 106      &  002.8+01.7 &            3.6 &            3.4 &             7 &  1.55 &   0.15 &  f656n \\
% 113      &  003.1+02.9 &            8.6 &            8.2 &           -23 &  4.29 &   0.31 &  f656n \\
% 114      &  003.1+03.4 &            3.4 &            3.4 &            74 &  1.45 &   0.31 &  f656n \\
% 139      &  004.0-03.0 &            3.4 &            3.3 &            70 &  3.41 &   0.35 &  f656n \\
% 151      &  004.8+02.0 &            3.6 &            3.6 &            42 &  0.70 &   0.08 &  f656n \\
...      &  ... &  ... &    ... & ... & ... & ... & ... \\
\bottomrule
\end{tabular}
\end{table*}

\section{Data Sets}

\subsection{Radio-Continuum Flux Densities}

Radio-continuum flux densities used for SED fitting are compiled from the literature \citep{\RadioReferences}. 

We considered only Galactic PNe (GPNe) designated as 'True' in the HASH catalogue \citep{Frew2010a}. The sample is furthermore filtered to GPNe with a missing value for the angular diameter or known angular diameter limited to 50~arcsec. The limit is chosen based on beam sizes of the two largest data sources in our sample (NVSS and SUMSS) and it was mainly applied to avoid, as much as possible, the common problem of missing flux \citep{Cornwell1988,pety2010}. Finally, out of \NoWithRadio\ radio detected GPNe we selected \NoFoundObjects\ objects with at least 3 data points in different radio bands for SED fitting.

\subsection{Angular Diameters Obtained in Conventional Methods}
 \label{sec:angdiams}

In this study we used two catalogues of angular diameters, one of diameters determined in radio bands and one in optical bands, in order to examine the accuracy of our models. 

The first catalogue is a compiled set of radio angular diameters collected from \cite{Seaquist1983, Phillips1988, Zijlstra1989a, Aaquist1990, Bains2003}. The methods used for determination of angular diameters in these papers were 10\%, Gaussian deconvolution and average visibility method. We note that the average visibility method is exclusively used for some of angular diameters from \cite{Aaquist1990}. However, since the exact method is not specified in the original paper, we will assume that all diameters smaller than 3~arcsec, from this paper, have been estimated using average visibilities.

We supplemented this set with new measurements for \NoNewVLADiams\ PNe found in National Radio Astronomy Observatory (NRAO) {\it Very Large Array} (VLA) Archive Survey (NVAS\footnote{\hyperlink{http://archive.nrao.edu/nvas/}{archive.nrao.edu/nvas}}) and not available in the examined literature. NVAS is an online database of VLA images automatically processed from a subset of the raw, non-proprietary uvdata in the VLA archive by the AIPS VLA pipeline (by Lorant Sjouwerman, NRAO). We explain the measurement process and results for this supplement in \cref{sec:angdiamnvas}. We will refer to this catalogue of diameters (including our new measurements) as Radio Literature (RL) diameters.

\begin{figure*}
  \centering
    \raisebox{0.09\height}{\includegraphics[width=0.288\textwidth]{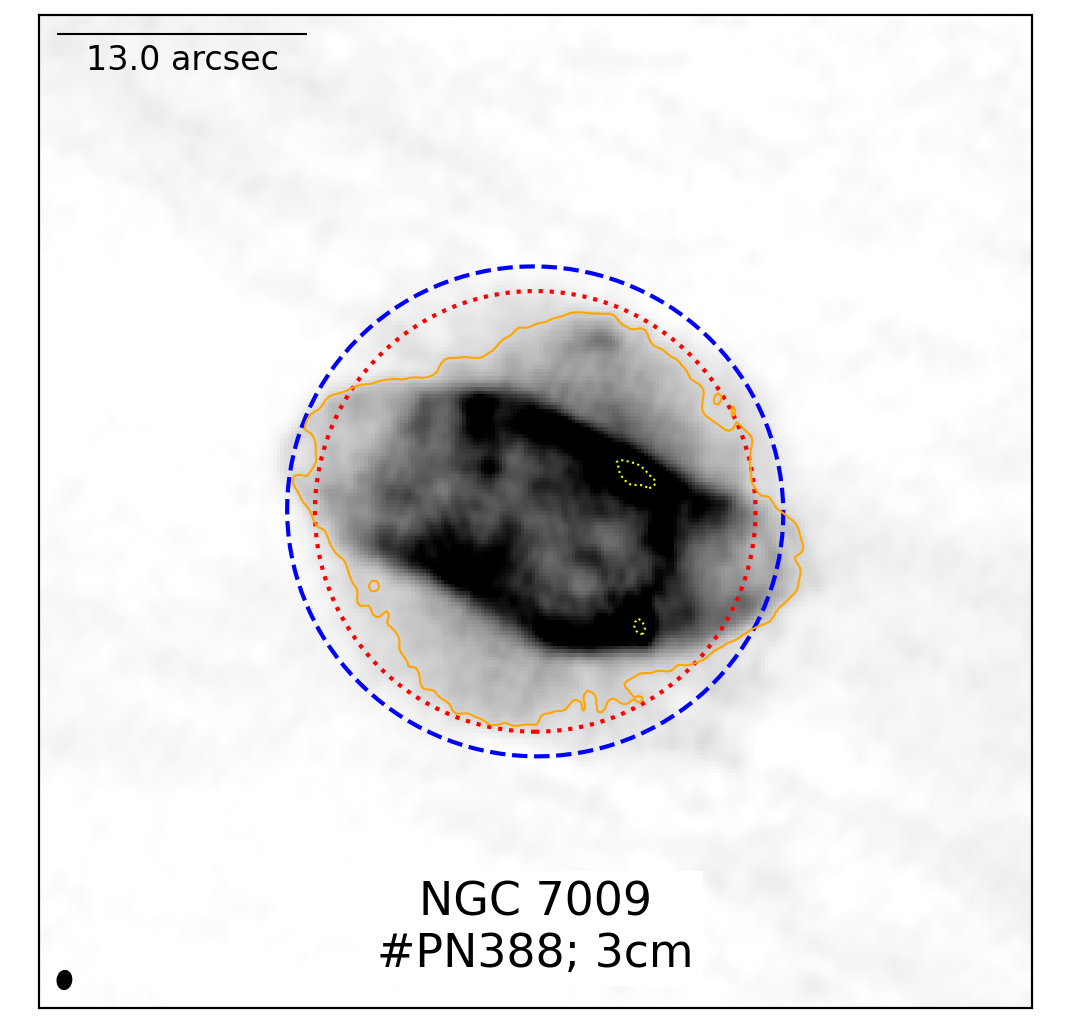}}
    % \hspace*{.2in}
    \raisebox{0.09\height}{\includegraphics[width=0.288\textwidth]{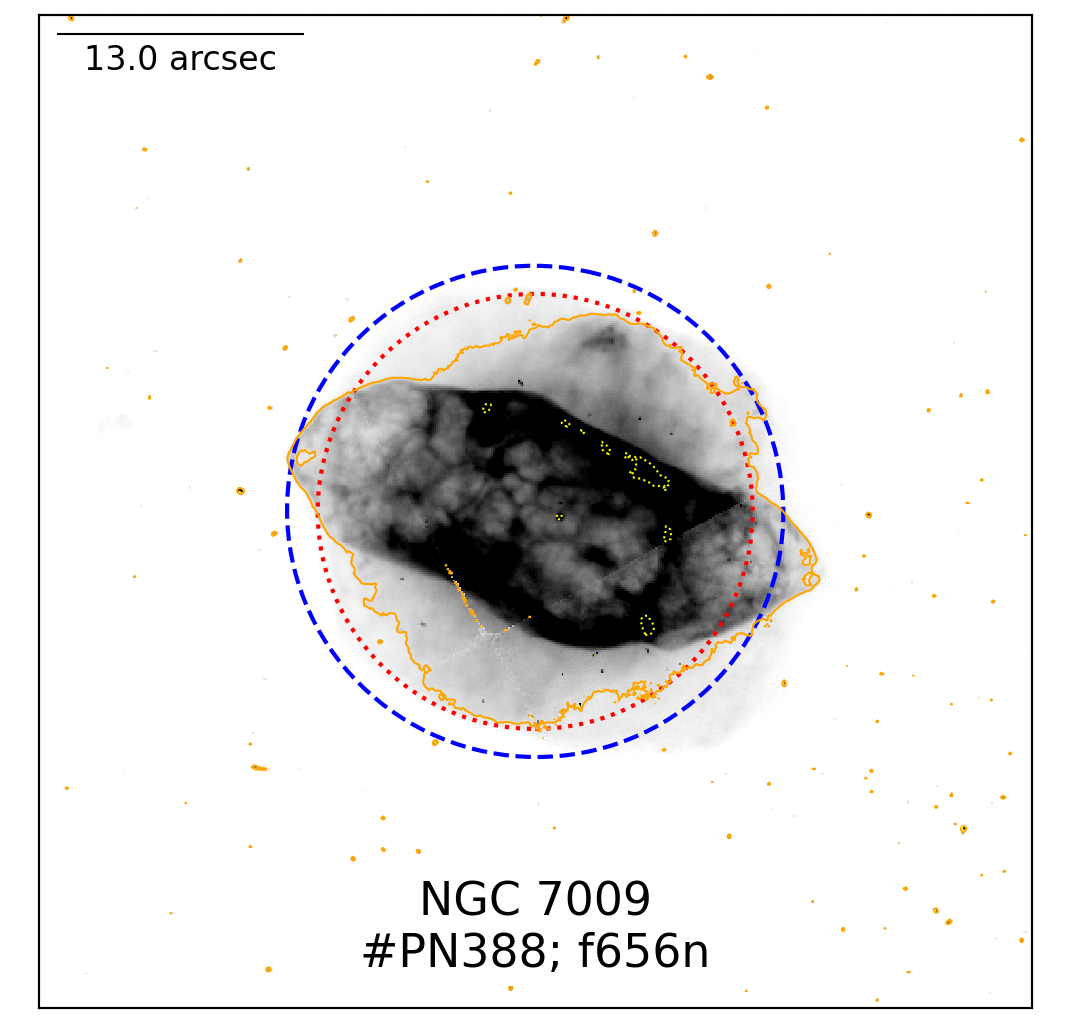}}
    % \hspace*{.2in}
    \raisebox{-0.0\height}{\includegraphics[width=0.39\textwidth]{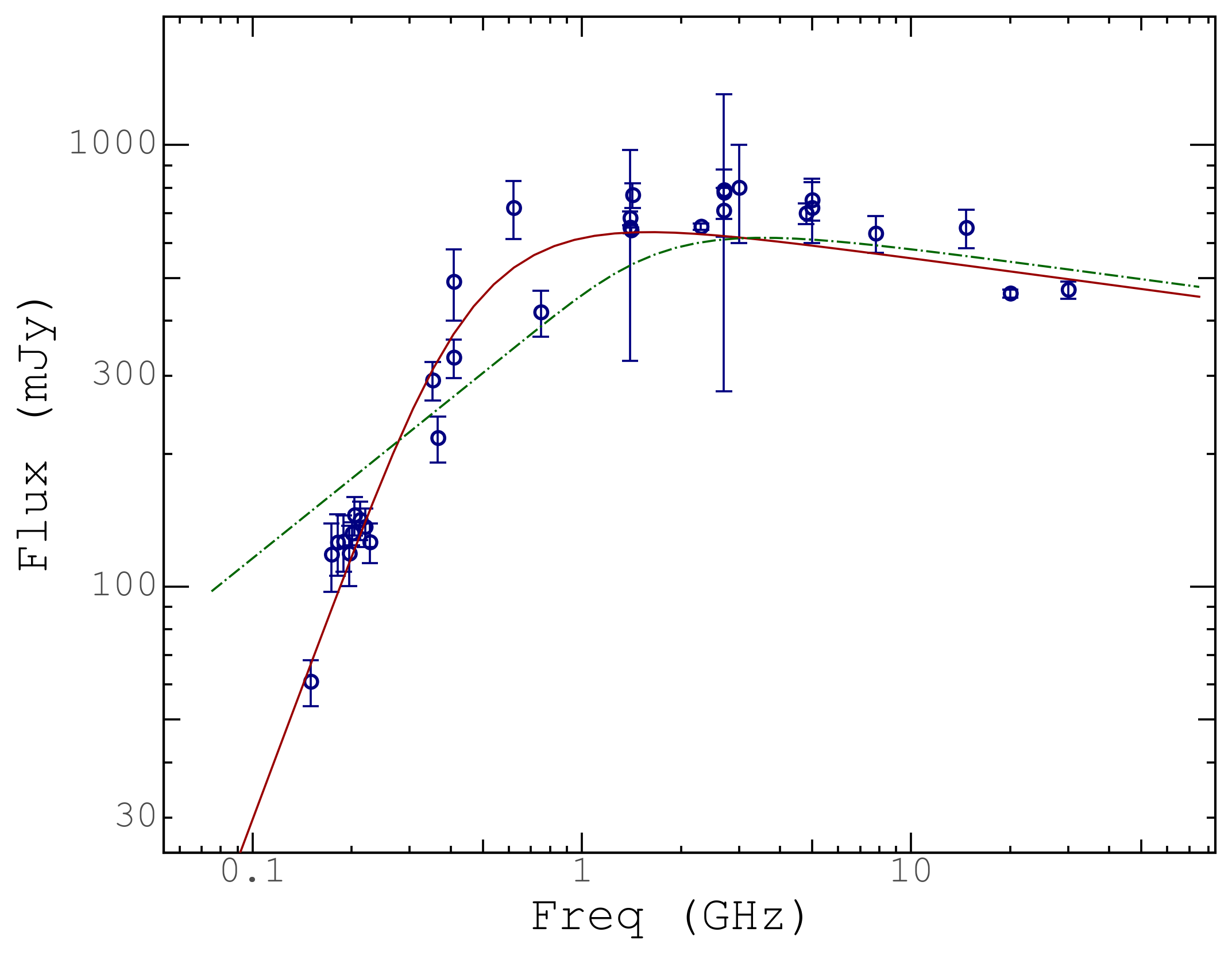}}
\caption{An example of angular diameter estimates presented in this paper. {\bf Left} and {\bf Middle:} VLA and HST images, respectively, overlaid with: angular diameter from 10\% method (blue dashed line), angular diameter from the SED method (red dotted line), peak emission (yellow dotted line) and 10\% level of the measured peak (orange full line).
{\bf Right:} Observed spectral energy distributions and best fits from \plshell\ (green, dash-dotted line) and \spshell\ (red, solid line) model. Results presented here are for PN NGC~7009. Full results are available as supplementary documents: AppendixA.pdf, AppendixB.pdf and AppendixC.pdf.}
 \label{fig:example_plots}
\end{figure*}

For the optical catalogue we measured optical angular diameters (using the 10\% method) of \NoNewHSTDiams\ PNe in our initial sample from the Hubble Legacy Archive (HLA\footnote{\hyperlink{https://hla.stsci.edu/}{hla.stsci.edu}}). We explain the measurement process and results for this dataset in \cref{sec:angdiamnvas}. This catalogue is supplemented with published values from \cite{Tylenda2003} and \cite{Ruffle2004}. We will refer to this catalogue as Optical Literature (OL) diameters.

\subsubsection{Newly measured radio angular diameters from NVAS and HLA images}
\label{sec:angdiamnvas}

We utilised a feature of the HASH PN database of providing a direct link to NVAS and HLA for PNe with available observations to obtain and examine high resolution VLA radio and multi-waveband {\it Hubble Space Telescope} (HST) images. 

We used {\sc Astropy's} \textsf{photutils.isophote} package to estimate major and minor axis and positional angle. All bright and compact background objects (including PN central stars) have been masked prior to fitting. Since the \textsf{isophote} method expects centre filled objects, for all PNe with a clear shell-like morphology, we adjust the starting position for fitting to the manually estimated ellipse placed on the brightest part of the nebula. The measurement method, for both sets, was that of 10\% peak brightness. For the peak brightness we used a level measured at the 99.5 percentile of the emission from the nebula.

For the VLA, we only selected images with a resolving beam smaller than the PN angular diameter quoted in the literature. We found angular diameter estimates for \NoNewVLADiams\ PNe from \NoFittedVLAImages\ radio images. These new measurements overlap with values found in the literature for \NoVLAOverlap\ PNe. For this cross-correlated sub-sample, except for six PNe, our estimates are in a good agreement with literature values. The six PNe with a large discrepancy ($>$30\% difference) are PNG~008.0+03.9 (\#PN:202), PNG~010.1+00.7 (\#PN:224), PNG~045.4--02.7 (\#PN:422), PNG~045.9--01.9 (\#PN:426), PNG~064.9--02.1 (\#PN:513) and PNG~161.2--14.8 (\#PN:695). We anticipate that the images we have used (produced by NVAS's automatic reduction pipeline) could be of lower quality than the ones used in the literature. Therefore, in the further analysis, we will use literature values for the cross-correlated sub-sample.

\begin{figure*}
 \includegraphics[width=\columnwidth]{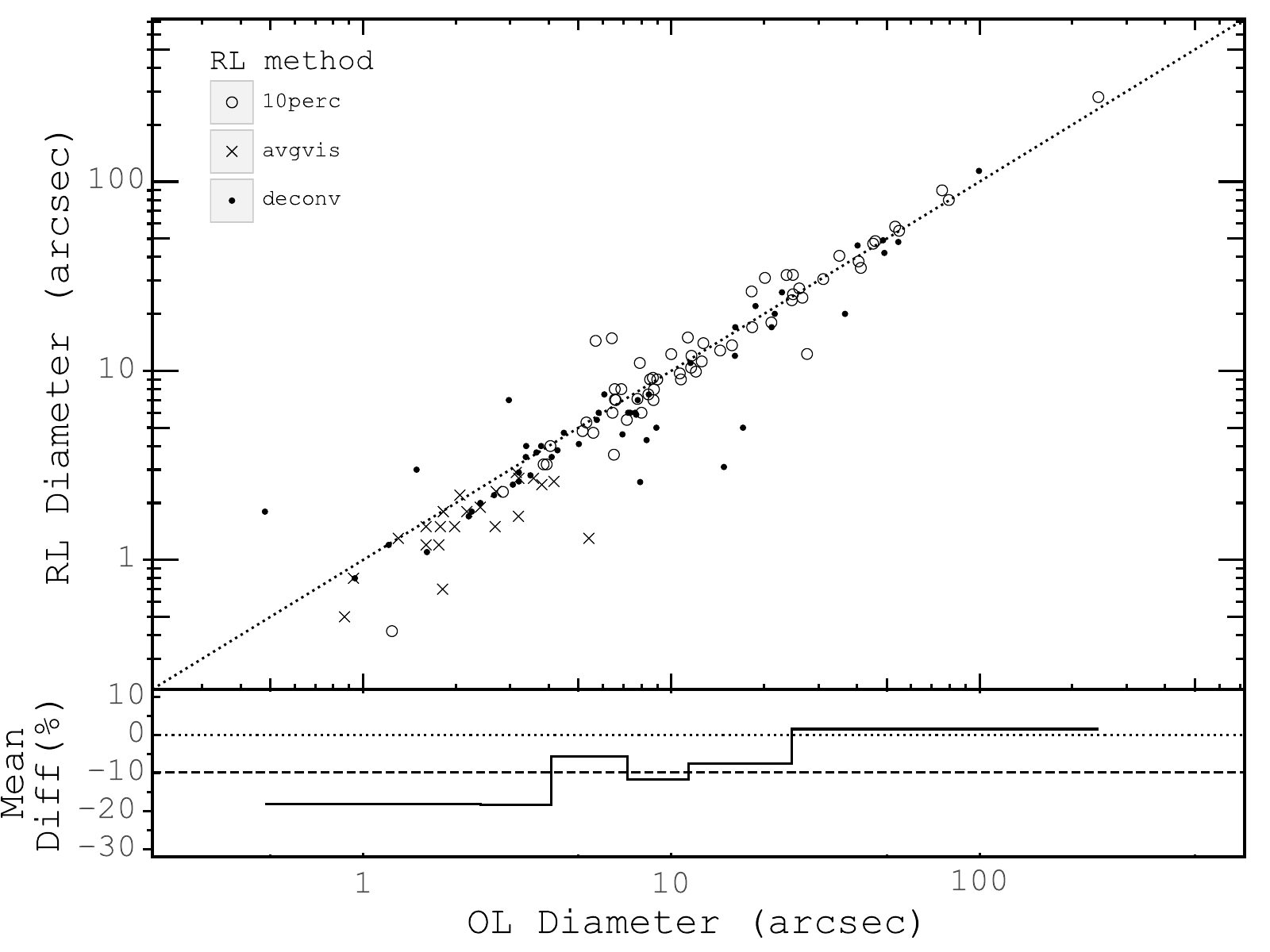}
 \includegraphics[height=6.32cm]{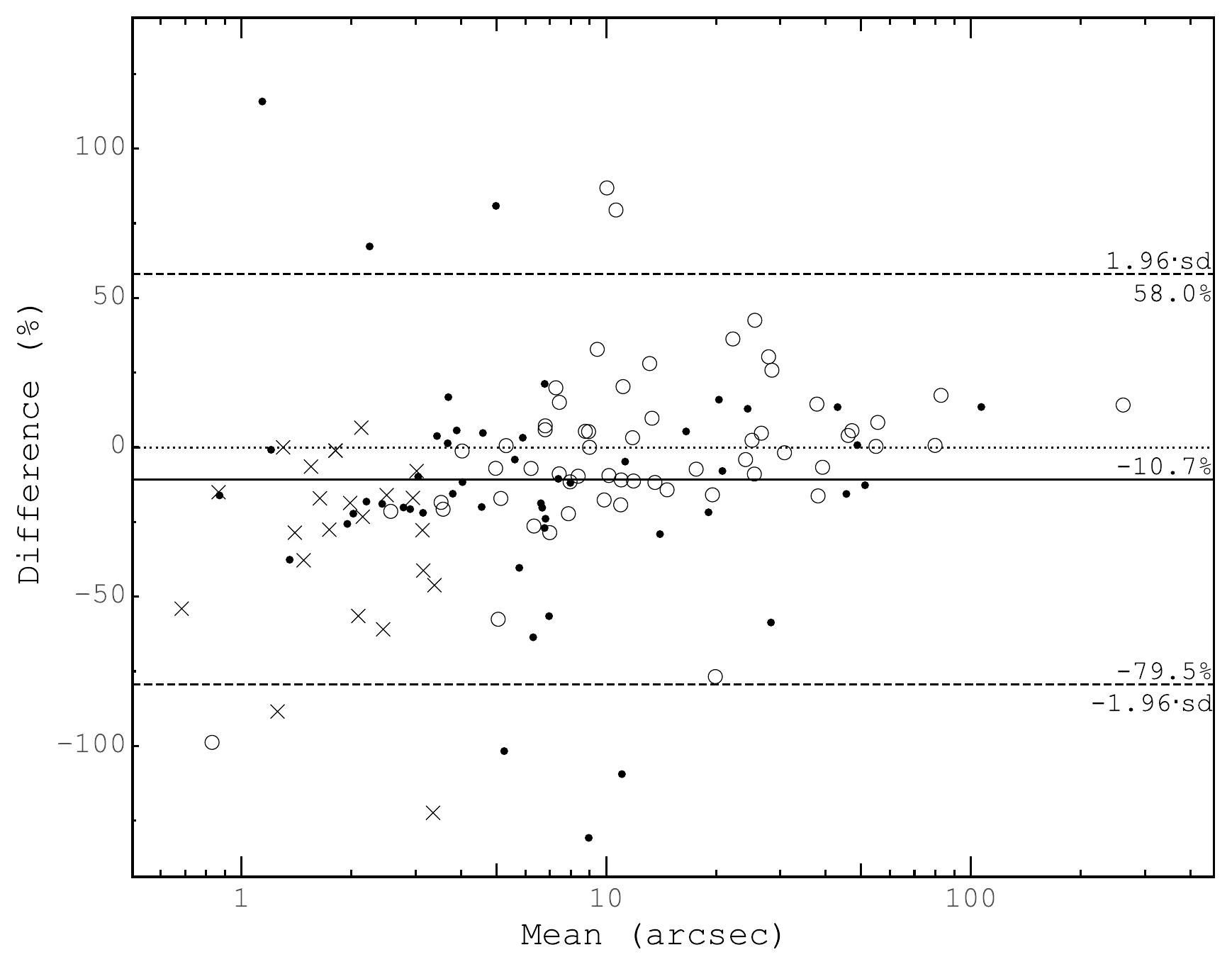}
 \caption{{\bf Left:} Direct comparison of angular diameters between OL (x axis) and RL catalogues (y axis). In {\bf bottom panel} we show medians of Mean Differences from same sample size bins. Dotted lines shows level on Mean Difference = 0 and dashed line shows the median of Mean Difference from the full sample. {\bf Right:} Tukey Mean Difference plot for comparison between OL and RL methods. Vertical lines represent: dashed line is 95\% limit of agreement between methods, solid line is the Mean Difference and the dotted line is zero level marker.}
 \label{fig:thetatheta_lit}
\end{figure*}

For the HST images we preferred H$\alpha$ (F656) images or, if not available, other commonly used optical filters. Our new measurements show excellent agreement ($<10\%$ difference) with the catalogue from \cite{Ruffle2004} (in further text R04). We noticed larger discrepancies in comparison with 10\% diameters from \cite{Tylenda2003} (in further text T03). T03 measured 10\% diameters from a large sample of GPNe using ground based observations in H$\alpha$ and H$\alpha$+[NII] from \cite{SCHWARZ1992} and \cite{Gorny1999}. For smaller PNe, of order of few arcsec, the HST has superior resolving power compared to capabilities of ground based telescopes. However, we found large discrepancies with T03 for 7 relatively large PNe as well: PNG~000.3+12.2 (\#PN:23), PNG~234.8+02.4 (\#PN:780), PNG~307.5--04.9 (\#PN:947), PNG~309.1--04.3 (\#PN:955), PNG~315.1--13.0 (\#PN:974), PNG~349.5+01.0 (\#PN:1144) and PNG~359.3--00.9 (\#PN:1309). Except for PNG~307.5--04.9, all objects in this group are bipolar nebulae with bright inner core. Our new 10\% diameters are mainly defined by the high brightness inner region with the much fainter bipolar lobes being below 10\% of the peak brightness. T03 noticed similar discrepancy in comparison with previous estimates from the literature but claimed that their estimates (in these cases) "encompassed a significant portion of the outer bipolar structure". Since we measured angular diameters from images with significantly higher angular resolution and sensitivity than used in T03 we believe that our 10\% estimates are more accurate and we will use our new results in the following analysis. 

Finally, we present excerpts or our newly estimated angular diameters from VLA and HST images in Table~\ref{tab:vlafitteddiams_short} and Table~\ref{tab:hstfitteddiams_short}, respectively. Full tables are provided as supplementary material. Fig.~\ref{fig:example_plots} (left and middle) shows an example of total intensity images overlaid with estimated diameters from 10\% and SED methods (see further text for more information on the SED method). Full results are provided as supplementary documents (AppendixA.pdf and AppendixB.pdf). 

\subsubsection{Comparison between optical and radio angular diameters}

The results between different methods (and even within the same method applied on the data obtained with different observational characteristics) could significantly differ. While disagreeing on which method gives better results \cite{Stasinska1991a} and \cite{Pottasch1992} both found a significant discrepancy between angular diameters determined from optical and radio measurements. \cite{Bedding1994} re-investigated 21 compact (1-2~arcsec) Galactic Bulge PNe in H$\alpha$ and found good agreement between radio and new optical methods but large disagreement with old optically determined values concluding that some of the previously found discrepancies can clearly be accounted to poor, less sensitive measurements found in the literature. \cite{Tylenda2003} measured and discussed differences between angular diameters obtained from three generally used methods, direct measurements at the 10\% level of the peak surface brightness, Gaussian deconvolution and second-moment deconvolution. They found that the smallest discrepancy between the 10\% and Gaussian deconvolution methods is achieved in the case of ''compact but not too small nebulae``. As noted by the authors, the Gaussian deconvolution method, in this range, gives fairly reliable results but systematically underestimates (by about 10\%) the diameters compared to the 10\% contour measurements.

\begin{table*}
\centering
\caption{Excerpt of the table of newly estimated angular diameters from SED modelling using \spshell\ model. The full table is available online (Table3.vot). The columns are as follows: (1) HASH PN ID; (2) PNG designation; (3) and (4) centroid position of the nebula from HASH PN catalogue; (5) SED estimated angular diameter; (6) SED estimated critical frequency; (7) SED estimated emission measure; (8) number of data points (radio fluxes) used in SED fitting; (9) $\mathsf{nrmse}$ value; (10) total fit quality (as defined in the text).}
\label{tab:sedsphshell_short}
\begin{tabular}{rcccrrrrrr}
\toprule
(1) & (2) & (3) & (4) & (5) & (6) & (7) & (8) & (9) & (10) \\
\#PN &          PNG &      RA J2000 &    DEC J2000 &   $\theta_{\mathrm{erd}}$ &        \nucr\ &             EM & n$_{points}$ & $\mathsf{nrmse}$ & $Q_{tot}$ \\
 & & (deg) & (deg) & (arcsec) & (GHz) & $10^6$ (pc cm$^{-6}$) & & (\%) & \\

\midrule
  23 &   000.3+12.2 &  255.3899 &  -21.8258 &    7.8 $\pm$ 0.3 &  0.63 $\pm$ 0.03 &  1.2 $\pm$ 0.1 &           10 &             29.0 &         2 \\
  28 &   000.4+01.1 &  265.6046 &  -27.9266 &    1.9 $\pm$ 0.2 &    1.3 $\pm$ 0.1 &      5 $\pm$ 1 &            3 &              6.4 &         1 \\
  37 &  000.7--02.7 &  269.5399 &  -29.7389 &    2.6 $\pm$ 0.4 &    0.8 $\pm$ 0.2 &  1.9 $\pm$ 0.8 &            4 &             11.0 &         2 \\
  40 &   000.7+03.2 &  263.7280 &  -26.5991 &    2.3 $\pm$ 0.6 &    0.8 $\pm$ 0.2 &      2 $\pm$ 1 &            3 &             19.0 &         2 \\
  50 &  000.9--02.0 &  269.0115 &  -29.1878 &  0.70 $\pm$ 0.05 &    2.2 $\pm$ 0.2 &     16 $\pm$ 3 &            3 &             26.0 &         2 \\
%   51 &   001.0+01.9 &  +17:40:27.40 &  -27:01:02.0 &    2.3 $\pm$ 0.4 &    1.0 $\pm$ 0.2 &      3 $\pm$ 1 &            3 &             15.0 &         2 \\
%   68 &  001.5--06.7 &  +18:16:12.27 &  -30:52:07.8 &  1.51 $\pm$ 0.03 &  3.92 $\pm$ 0.09 &     54 $\pm$ 3 &           13 &              9.1 &         1 \\
%   69 &   001.6+01.5 &  +17:43:16.95 &  -26:44:17.6 &  1.17 $\pm$ 0.09 &    1.4 $\pm$ 0.1 &      7 $\pm$ 1 &            3 &             26.0 &         2 \\
%   74 &  001.7--04.4 &  +18:07:14.55 &  -29:41:24.4 &  0.45 $\pm$ 0.06 &    2.4 $\pm$ 0.4 &     20 $\pm$ 6 &            3 &             30.0 &         3 \\
%   81 &  002.0--06.2 &  +18:15:06.53 &  -30:15:32.9 &    2.5 $\pm$ 0.7 &    0.7 $\pm$ 0.2 &  1.3 $\pm$ 0.8 &            4 &             58.0 &         3 \\
  ... & ... & ... & ... & ... & ... & ... & ... & ... & ... \\
\bottomrule
\end{tabular}
\end{table*}

\begin{table*}
\centering
\caption{Excerpt of the table of newly estimated angular diameters from SED modelling using \plshell\ model. The full table is available online (Table4.vot). The columns are the same is in Tab.~\ref{tab:sedplshell_short}.}
\label{tab:sedplshell_short}
\begin{tabular}{rcccrrrrrr}
\toprule
(1) & (2) & (3) & (4) & (5) & (6) & (7) & (8) & (9) & (10) \\
\#PN &          PNG &      RA J2000 &    DEC J2000 &   $\theta_{\mathrm{erd}}$ &        \nucr\ &             EM & n$_{points}$ & $\mathsf{nrmse}$ & $Q_{tot}$ \\
 & & (deg) & (deg) & (arcsec) & (GHz) & $10^6$ (pc cm$^{-6}$) & & (\%) & \\

\midrule
23 & 000.3+12.2 & 255.3899 & -21.8258 & 3.82 $\pm$ 0.17 & 2.2 $\pm$ 0.2 & 16 $\pm$ 3 & 10 & 52.2 & 3\\     
48 & 000.9--04.8 & 271.7756 & -30.5714 & 4.9 $\pm$ 2.3 & 0.9 $\pm$ 0.8 & 2 $\pm$ 4 & 3 & 9.8 & 1\\
51 & 001.0+01.9 & 265.1142 & -27.0172 & 1.5 $\pm$ 0.2 & 2.5 $\pm$ 0.6 & 22 $\pm$ 11 & 3 & 0.07 & 1\\
68 & 001.5--06.7 & 274.0511 & -30.8688 & 0.37 $\pm$ 0.03 & 30 $\pm$ 5 & 3855 $\pm$ 1449 & 13 & 27.2 & 2\\
90 & 002.2--09.4 & 277.2985 & -31.4998 & 5.1 $\pm$ 0.7 & 1.1 $\pm$ 0.3 & 4 $\pm$ 2 & 3 & 35.8 & 3\\     
... & ... & ... & ... & ... & ... & ... & ... & ... & ... \\
\bottomrule
\end{tabular}
\end{table*}

While the aim of this paper is not a comprehensive comparison between radio and optical diameters we briefly examined available data from these two basic methods in order to set {\it a priori} precision limits that can be used for testing our new method in comparison with previous methods. In Figure~\ref{fig:thetatheta_lit} we show comparison between angular diameters from the optical and the radio catalogue. In the left panel we show a direct plot of RL diameters versus OL diameters and a step plot of Mean Differences averaged in same size bins. The Mean Differences (in per cents) is here defined as:

\begin{equation}
\textrm{Mean Diff} = 100\%\ \frac{(\theta_{\mathrm{RL}} - \theta_{\mathrm{OL}})}{0.5\ (\theta_{\mathrm{OL}}+\theta_{\mathrm{RL}})}
\end{equation}
i.e. the difference between angular diameters normalised to its mean. The two methods appear to be in relatively good agreement with a larger bias seen for smaller PNe ($<$5~arcsec) and no apparent bias for larger PN ($>$30~arcsec).

In the right panel we show Tukey Mean Difference plot (also called the Bland-Altman plot: \cite{bland1999}) for comparison between measurements obtained by different methods. In this plot the x-axis is the Mean Difference (as previously defined) and the y-axis is the mean between the two measurements. The bias for the full sample is $\sim$11\% i.e. the estimated radio diameters are, overall $\sim$11\% smaller than their optical counterparts. The larger disagreement appears to be for two indirect radio methods: Gaussian deconvolution and average visibilities. The same inconsistent bias, also visible in the left panel, towards smaller PNe, is also visible here. 

Since none of the methods used here can be considered as a ``gold standard'' it is hard to draw a decisive conclusion on which method is most responsible for the discrepancies. However, the best agreement with optical diameters is the 10\% radio method which shows the smallest bias of --1\% and smallest range at the 95\% confidence level of $\pm$55\%. Therefore, we accepted $\pm$55\% as the tolerable limit of disagreement between different angular diameter determination methods.

\section{Application of the SED Modelling}

We applied selected models to the sample of radio detected PNe from our base catalogue with data in at least three different radio bands. The two presented models have been applied with \nucr\ and $\theta$ left as free parameters (i.e the emission measure through the centre of the nebula and the outer angular diameter of the shell, respectively). If the standard error of the fit, calculated from the obtained covariance matrix, was larger than 100\% for any of the fitted parameters, we rejected the fit. Also, the fit was rejected if no data point was available in the optically thick part, i.e. if the minimal frequency in the fitted data set was higher than arbitrary selected limit of 1.5 $\cdot$\nucr\. Consequently, about 30\% of the initial sample was rejected and was not used in the final analysis.

\subsection{Assessing the Fit Quality}

In principle, the estimate of angular diameters can be established from the SED fit based only on two points, as long as one flux measurement is from the optically thick and one from the optically thin part of the radio-continuum spectrum. However, in such a case it would be hard to measure fit quality and establish a confidence level of the final result. Clearly, the final estimates will depend not just on the used model and validity of the used approximations but also on the SED data coverage i.e. on the number and the accuracy of the data points used in the fit. 

We estimated a rough quality of the fit using normalised root mean square residuals ($\nrmse$) in combination with consideration of the number of available data points in the optically thick and optically thin regions. Using $\nrmse$ alone could pose a problem with very small data-sets. For example, for a minimal data set (e.g. 3 data points), the model could have a very small $\nrmse$ however, the solution will strongly rely on only one point in the optically thick or optically thin region. Normalisation of $\nrmse$ for each fit is done using the estimated standard deviation in order to place all $\nrmse$ values on a similar scale i.e. the $\nrmse$ is calculated as:

\begin{equation}
    \nrmse = 100\%\ \frac{1}{\sigma(S)}\ \sqrt{\frac{\sum_{i=1}^{n}(S_i-\hat S_i)^2}{n}}
\end{equation}

where $S_i$ and $\hat S_i$ are measured and predicted flux density for i-th observation, $\sigma(S)$ is the standard deviation of measured flux densities and $n$ is the number of observations.

Therefore, in order to make a distinction between the quality of SED fits, we adopted the following scheme:

\begin{figure*}
 \includegraphics[width=\columnwidth]{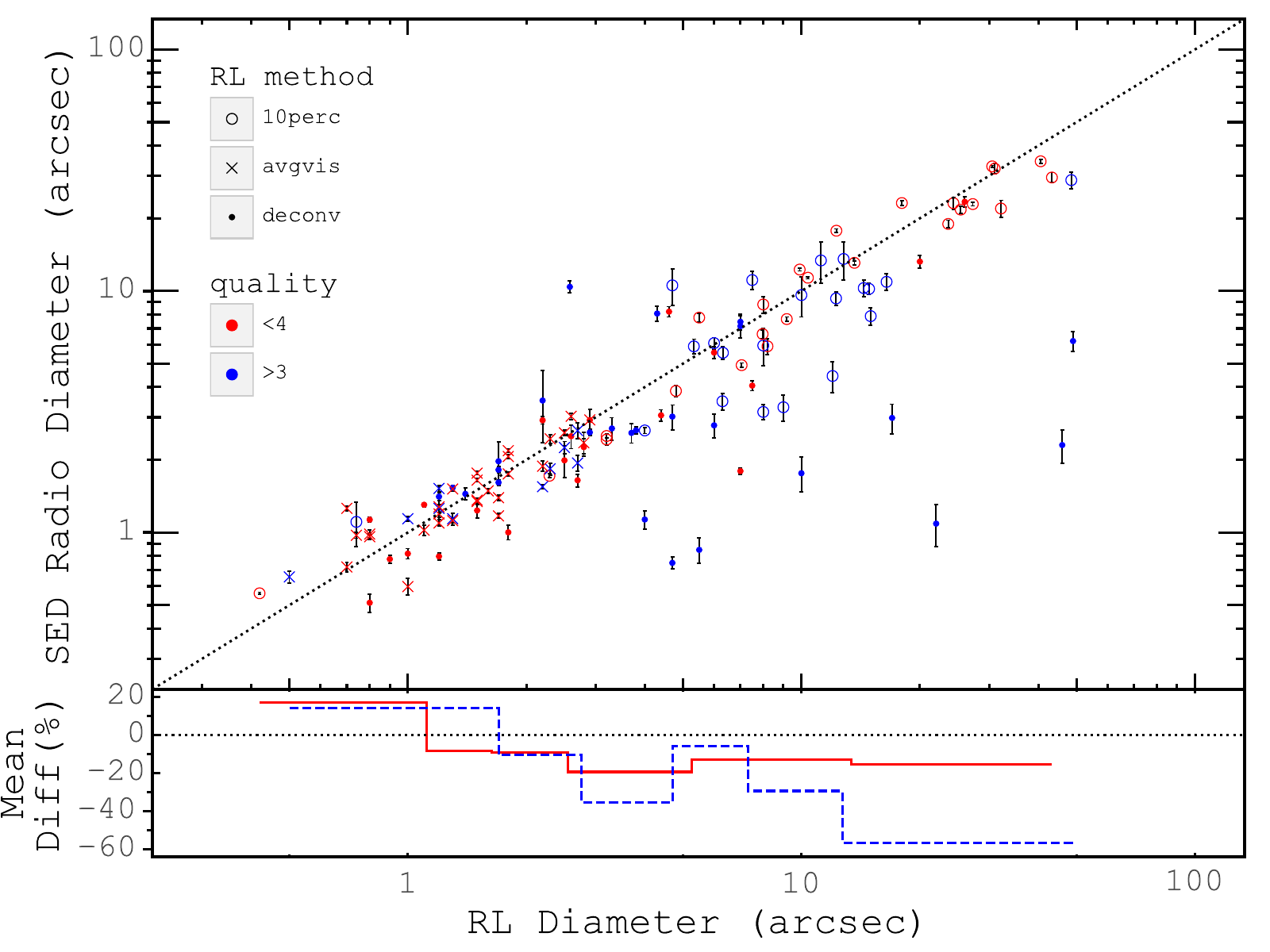}
 \includegraphics[width=\columnwidth]{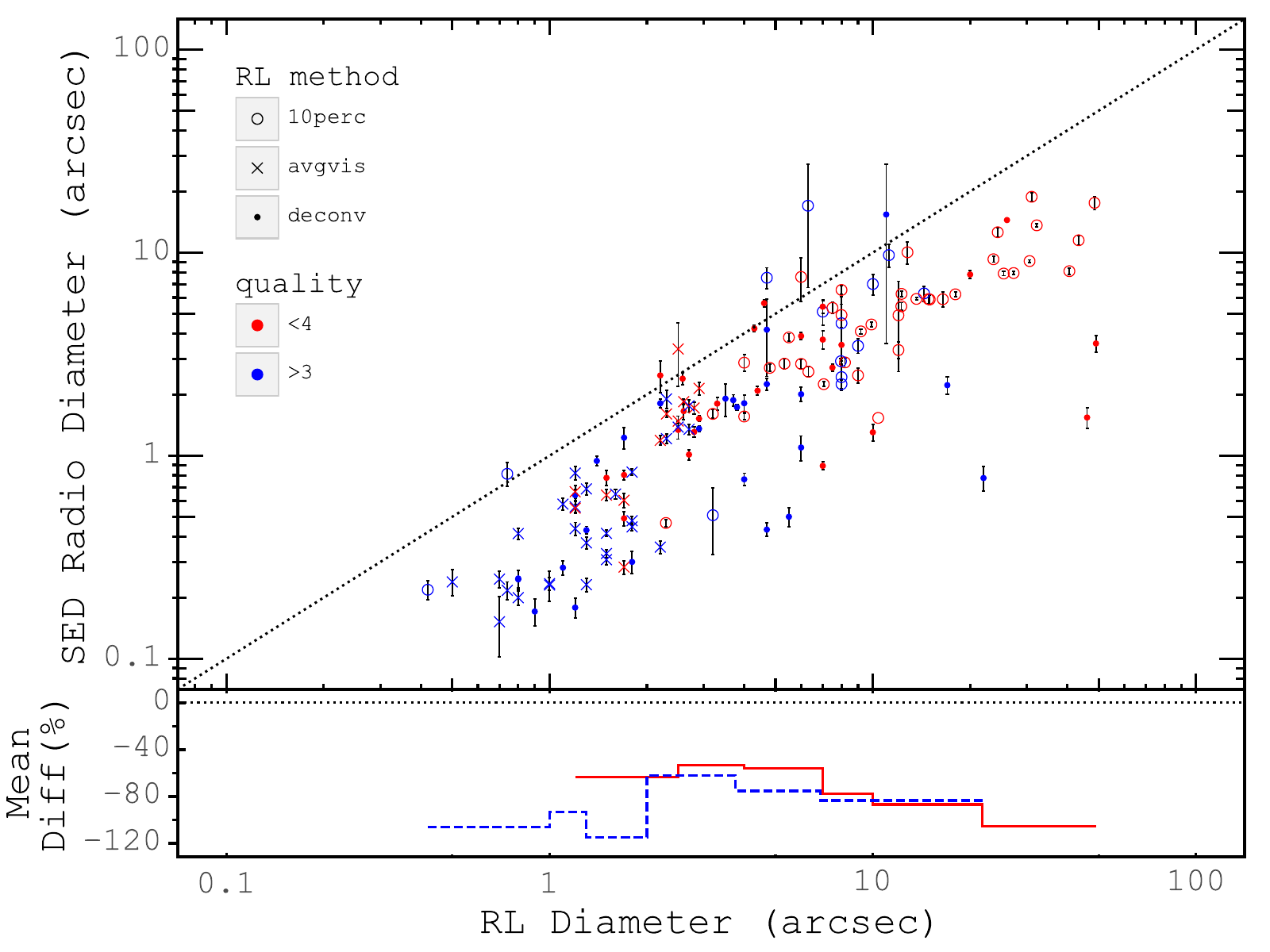}
  \includegraphics[width=\columnwidth]{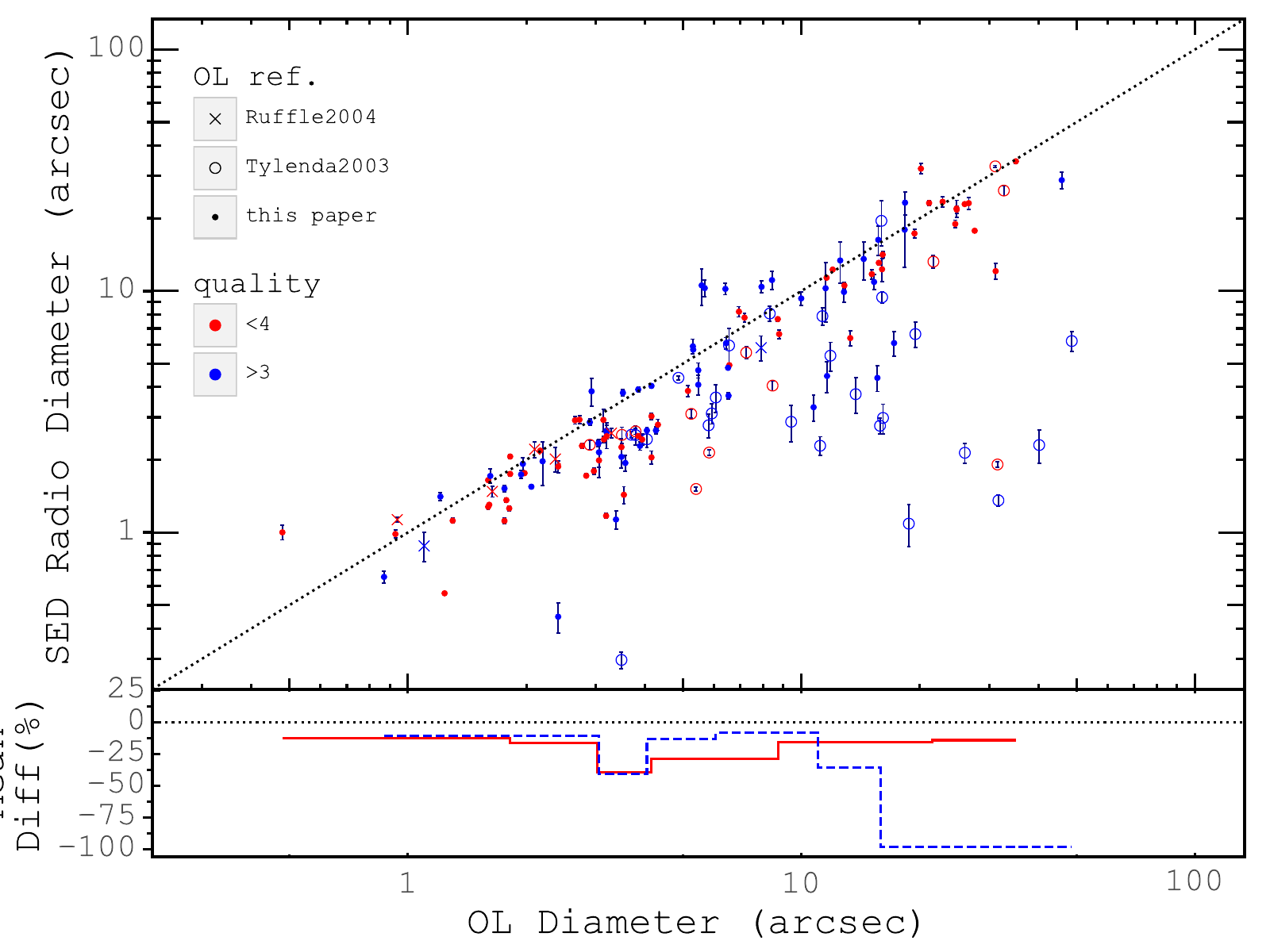}
 \includegraphics[width=\columnwidth]{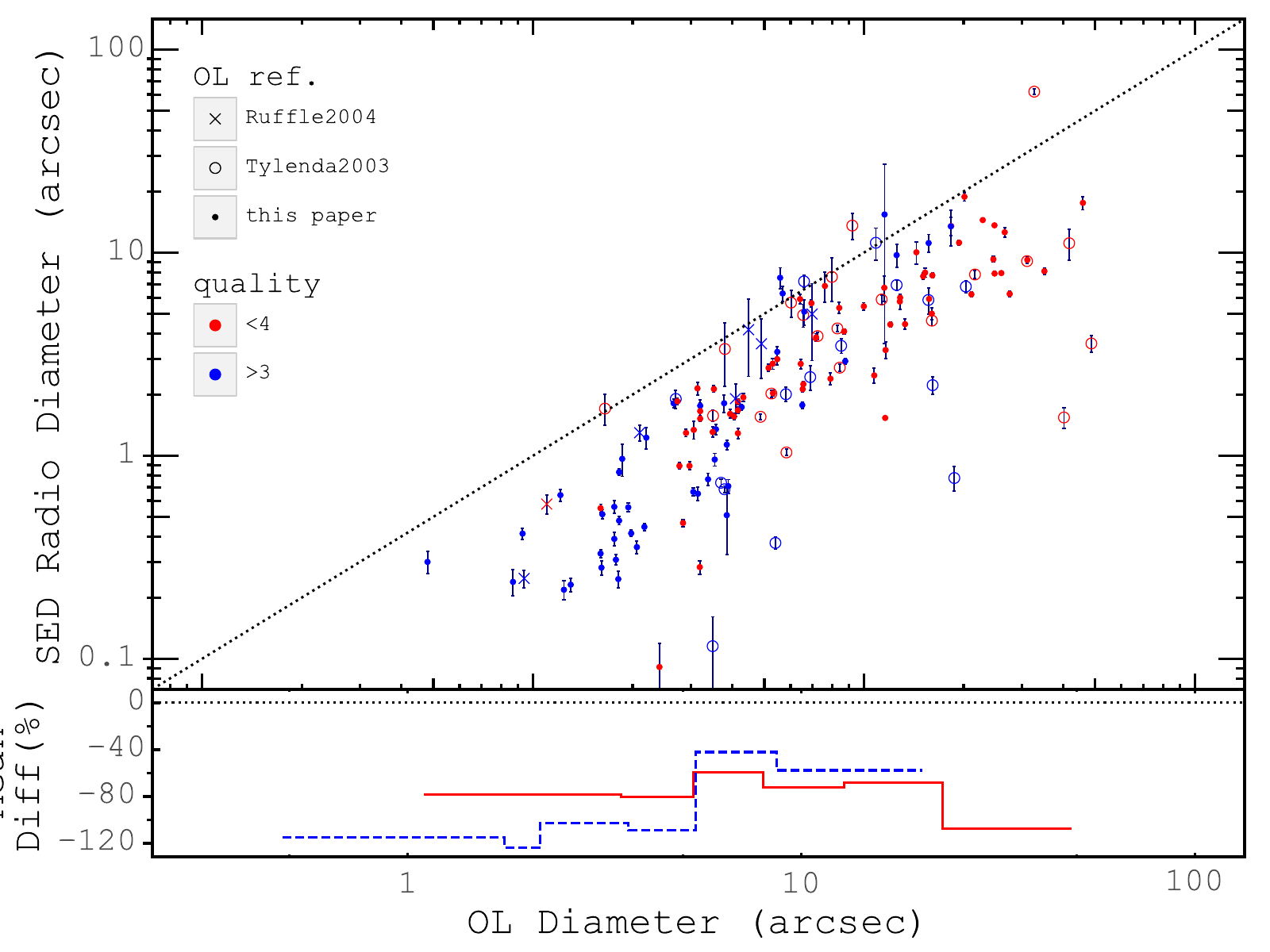}
 \caption{Direct comparison of angular diameters estimated in our models in comparison with RL ({\bf top row}) and OL ({]bf bottom row}) estimates. {\bf Top panels:} The point shapes used indicate different RL/OL methods and the dotted line is the 1-1 relationship. Red and blue data points represent SED diameter estimates with total fit quality $Q_{tot}\leq3$ and $Q_{tot}>3$, respectively. {\bf Bottom pannels:} Same as in Figure~\ref{fig:thetatheta_lit} where the red solid line and blue dashed line represent binned Mean Difference medians of SED diameter estimates with total fit quality $Q_{tot}\leq3$ and $Q_{tot}>3$, respectively. {\bf Left column:} {\it sp shell} SED model and {\bf right column:} {\it pl shell} SED model. Note that correction of 1.84 is applied to values estimated from \plshell\ model (see \cref{sec:plshell} for more details).}
 \label{fig:thetathetashells}
\end{figure*}

\begin{enumerate}
    \item First, we determined $\nrmse$ quality ($Q_{\nrmse}$) by applying hard limits to $\nrmse$ values where:
    \begin{equation}
        Q_{\nrmse} = 
            \begin{cases}
                1, \text{ for } \nrmse < 10\%\\
                2, \text{ for } 10\% < \nrmse < 30\% \\
                3 , \text{ for } \nrmse > 30\%
            \end{cases}
    \end{equation}
    \item Next, we determined a quality factor based on number of data points in two ranges of optical thickness ($Q_{\tau}$):
    \begin{equation}
        Q_{tau} = 
            \begin{cases}
                1, \text{ for at least 2 points in both regions}\\
                2, \text{ for at least 2 points in optically thick} \\
                \text{\qquad and only 1 point in optically thin region }\\
                3 , \text{ for only 1 point in optically thick region}
            \end{cases}
    \end{equation}    
    \item The final quality factor is calculated as $Q_{tot}=Q_{\nrmse}\cdot Q_{\tau}$
\end{enumerate}

\subsection{Fit Results}

We attempted SED fits for \NoFoundObjects\ PNe. We successfully estimated angular diameters and EM for the \NoSEDfittedspshell\ and \NoSEDfittedplshell\ PNe in \spshell\ and \plshell\ models, respectively. Full tables with fitting results from both models are accessible as supplementary material. We show examples of tables layouts in Table~\ref{tab:sedsphshell_short} and Table~\ref{tab:sedplshell_short}. 

We provide SED plots of all, successfully fitted, models in a supplementary document (AppendixC.pdf). As an example we present estimated SED models for PN NGC~7009 in Fig.~\ref{fig:example_plots} (right).

\subsection{Comparison with Other Methods}

\begin{figure*}
 \includegraphics[width=\columnwidth]{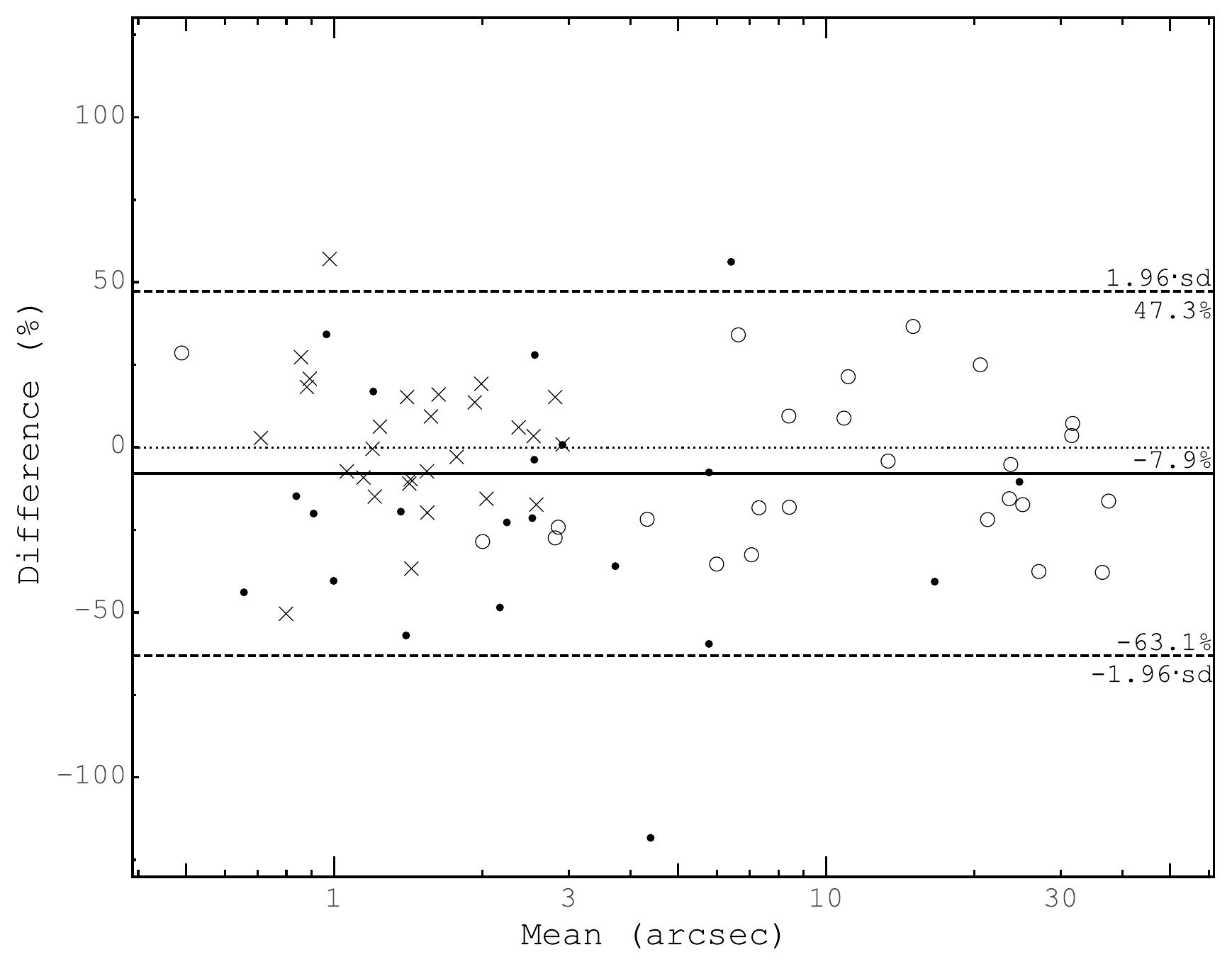}
 \includegraphics[width=\columnwidth]{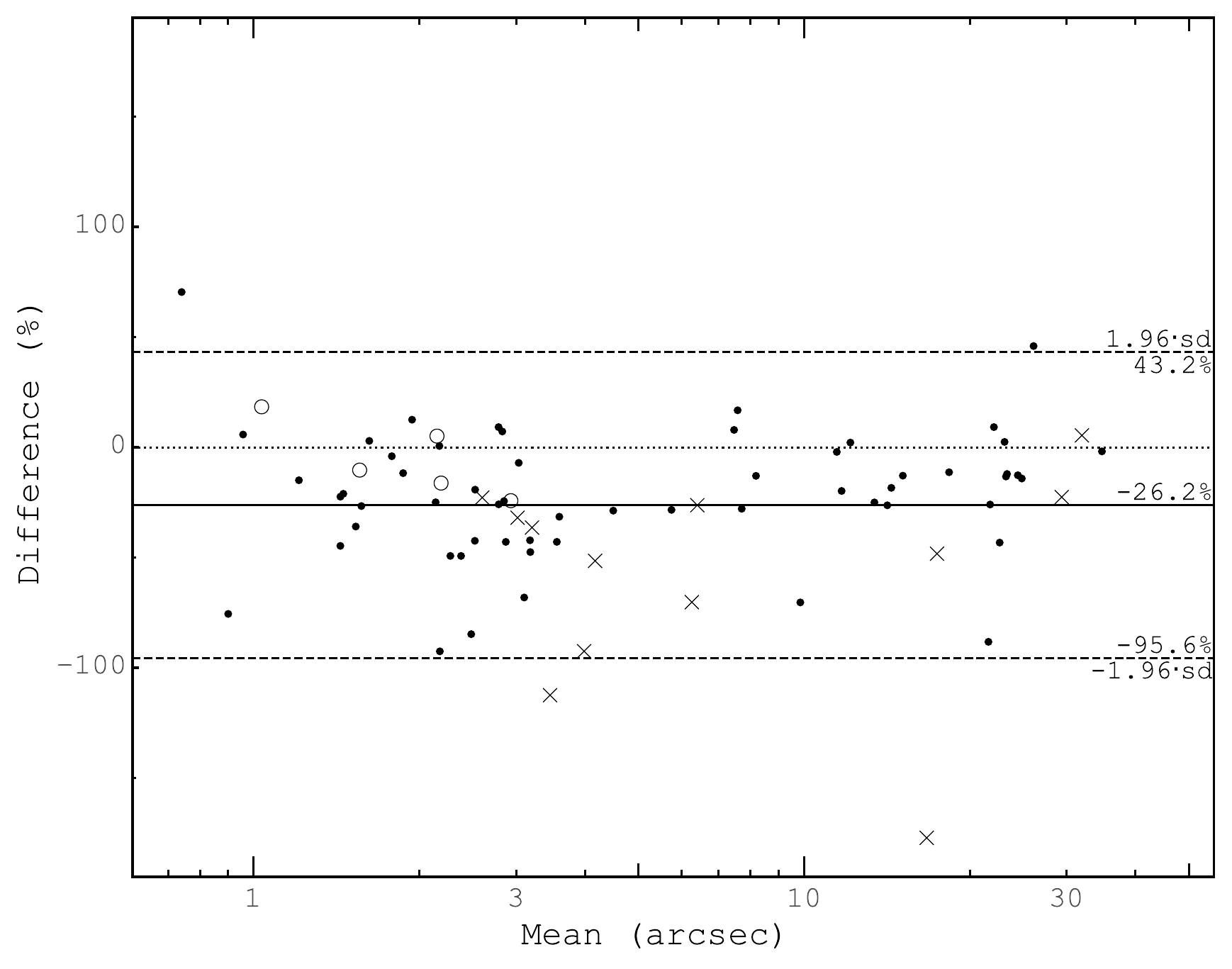}
 \includegraphics[width=\columnwidth]{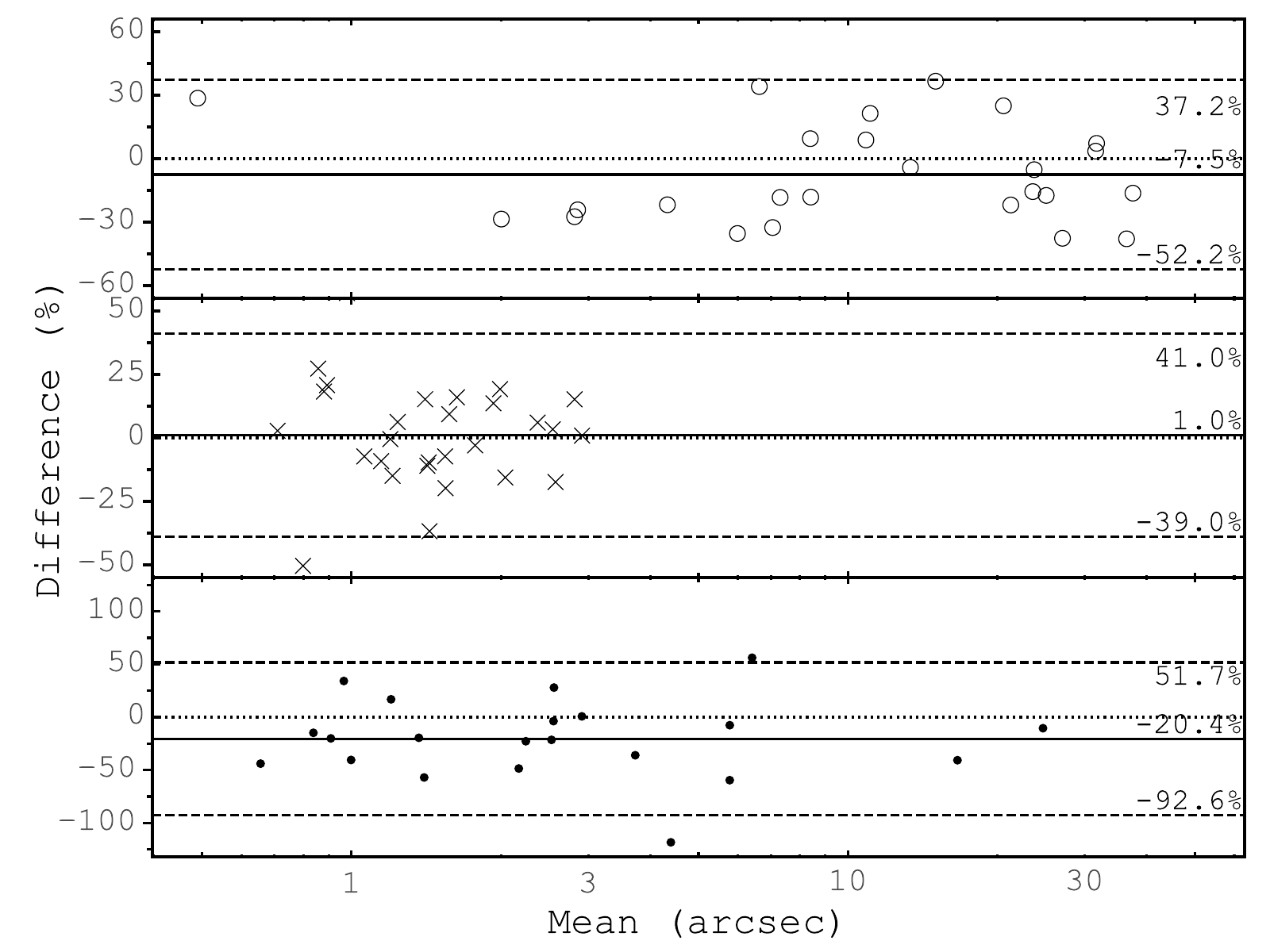}
 \includegraphics[width=\columnwidth]{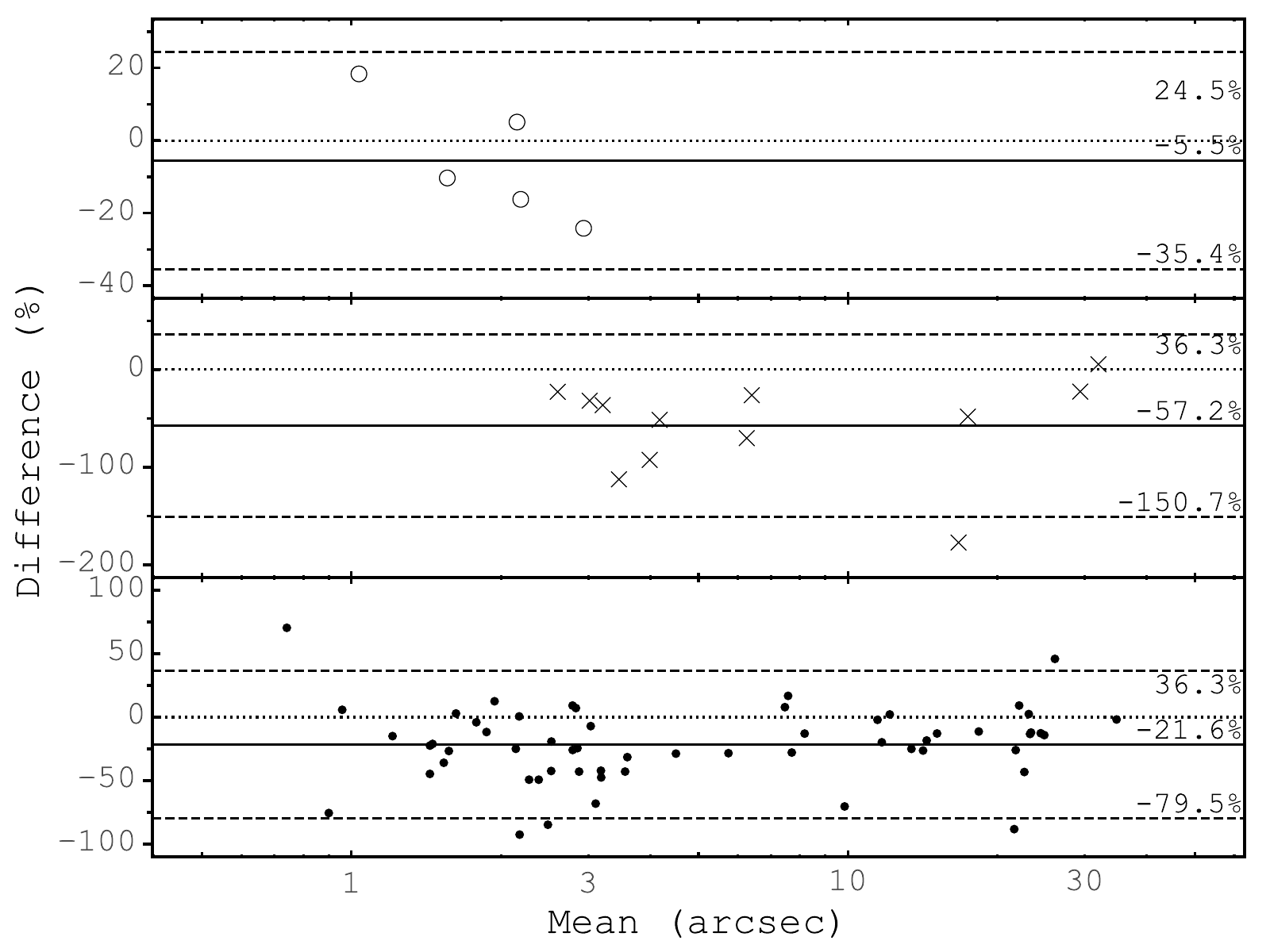}
 \caption{Tukey Mean Difference plot for comparison between results from our method and methods found in the literature. The vertical lines in all plots represent: dashed line is 95\% limit of agreement between methods, solid line is the Mean Difference and the dotted line is zero level marker. {\bf Top left:} combined results from all methods in comparison to the RL catalogue, the Mean Difference is calculated as Mean Difference = 100\% $\cdot(\theta_{\mathrm{erd}} - \theta_{\mathrm{RL}})/\overline{(\theta_{\mathrm{erd}},\theta_{\mathrm{RL}})}$. {\bf Top right:} combined results from OL catalogue, Mean Difference = 100\% $\cdot(\theta_{\mathrm{erd}} - \theta_{\mathrm{OL}})/\overline{(\theta_{\mathrm{OL}},\theta_{\mathrm{erd}})}$. {\bf Bottom row, left:} comparison by RL catalogue method and {\bf right} comparison by OL reference, the Mean Differences are calculated in the same way as in the top row plots. Note that the subplots in the bottom row are presented in different ranges.}
 \label{fig:tukeyplots}
\end{figure*}

In Figure~\ref{fig:thetathetashells} we show direct comparison plots between our models and RL and OL estimates. We found \noRLandspshelldiams\ and \noRLandplshelldiams\ counterparts in the RL catalogue and \noOLandspshelldiams\ and \noOLandplshelldiams\ in the OL catalogue for \spshell\ and \plshell\, respectively. The methods used in the RL catalogue are indicated with open circles, x-es and filled circles for 10\%, average visibility and deconvolution, respectively. Similarly, three base OL catalogues, R04, T03 and HST catalogue from this paper were indicated with x's, open circles and filled circles, respectively. The scatter plots were colour coded to show the overall fit quality as red for good quality ($Q_{tot}\leq3$) and blue for poor quality fits ($Q_{tot}>3$). 

While the results from the \spshell\ model show an overall agreement with both RL and OL values, the \plshell\ model, clearly, systematically underestimates diameters in comparison with both catalogues. This result from the \plshell\ model was generally expected since, as we stated in \cref{sec:plshell}, the optically thick component in this model has a slope with an upper limit of possible values for a nebula with total emission affected by both shell and slow expanding and undisturbed AGB wind. This result confirms findings from \cite{Hajduk2018} where the authors used several comprehensive SED models to construct and examine a large set of PNe radio SEDs. The results in the aforementioned paper could not confirm a strong radial density gradient in any of the examined objects, which are to be expected in our \plshell\ model. However, by examining modelled spectra in more detail we found 3 PNe for which constructed SEDs show some evidence of shallower spectra in the optically thick region than as expected from a constant density, spherical shell. For this test we only examined objects with at least 3 data points in the optically thick and at least 2 points in the optically thin part of the spectrum. We found 3 objects which appear to have better agreement (lower $\nrmse$) with \plshell\ model: PNG~312.3+10.5 (\#PN:965), PNG~058.3--10.9 (\#PN:483) and PNG~002.0--13.4 (\#PN:4103).

Another important result from \cite{Hajduk2018} is that the usual, homogeneous density models (similar to approximations used in this paper) cannot fully explain the SEDs of most PNe.  Although, in principle, we do agree with this finding, including extra free parameters in our model (e.g. radially dependent density distribution or sky projected geometry) would render our method impractical to use. 

Secondly, as expected, good quality fits show better and more consistent agreement with RL and OL values. In comparison with RL values, the distribution of Mean Difference medians for good quality factors ($Q_{tot}\leq4$) appear to be within 20\% from the ideal agreement. On the other hand, the poor quality fits show significantly larger disagreement, especially for PNe with angular diameters larger than $\approx10$~arcsec. A similar trend is seen in comparison with OL values but with an additional bias of $\approx$20\%. 

Since we define the total quality as a combination of residuals size and the SED data coverage, low total quality does not necessarily mean inaccuracy of the final estimate. Some of the fitting results will score low in $Q_{tot}$ only because of the low data coverage while the estimated angular diameter could actually be relatively accurate (as the majority of low quality estimates appear to be quite close to the 1--1 reference line). 

Finally, we conclude that the \plshell\ model cannot sufficiently well explain the observed SEDs from any PNe in our sample and we will not use it in the further analysis. Also, in order to avoid large bias inherited from low quality data, we will use only results with $Q_{tot}\leq3$ for the comparison with previous methods.

In Figure~\ref{fig:tukeyplots} we show Tukey Mean Difference plots for comparison between our new method and the OL and RL catalogues. We show combined results (top row), and results separated by the methods used in the literature. Dashed lines in all plots show the 95\% level of confidence agreement within two compared methods and solid lines show the mean of Mean Difference.

As can be seen from presented plots, the SED method is almost always negatively biased in comparison with other methods i.e. it almost always results in smaller estimates. In comparison with all RL values the bias is $\approx$9\% with 95\% confidence within the range of $\pm$55\% around the central tendency. The comparison with all OL values shows significantly larger bias ($\approx$27\%) and the larger span of values within the 95\% confidence range (of $\pm$70\%). In comparison with separate RL methods it can be seen that the agreement is significantly better for 10\% and average visibilities method. The overall discrepancy in both bias and confidence range appear to be mainly affected by differences with the deconvolution method.

Similarly, discrepancies between SED and OL values are mainly arising from comparison with T03 estimates. In comparison with our new HST diameters, the SED values show $\approx$--20\% bias and 95\% confidence range, around central tendency, of about $\pm60$\% in Mean Difference.

Except for the larger bias, the results are similar to the results of comparison between OL and RL methods. The range of Mean Difference is very close to the acceptable value adopted from OL vs. RL angular diameters consideration. That implies that the range of intrinsic (unmeasurable) uncertainties, arising from the method itself, is similar to previously used methods. The bias is higher and very likely inherent from the approximations used in the method. For the values presented in this paper we propose a bias correction of 15\% which should place SED angular diameters between OL and RL estimates (note: this correction is not applied to current values).

\section{Auxiliary results: Emission Measure}

\begin{figure}
 \includegraphics[width=\columnwidth]{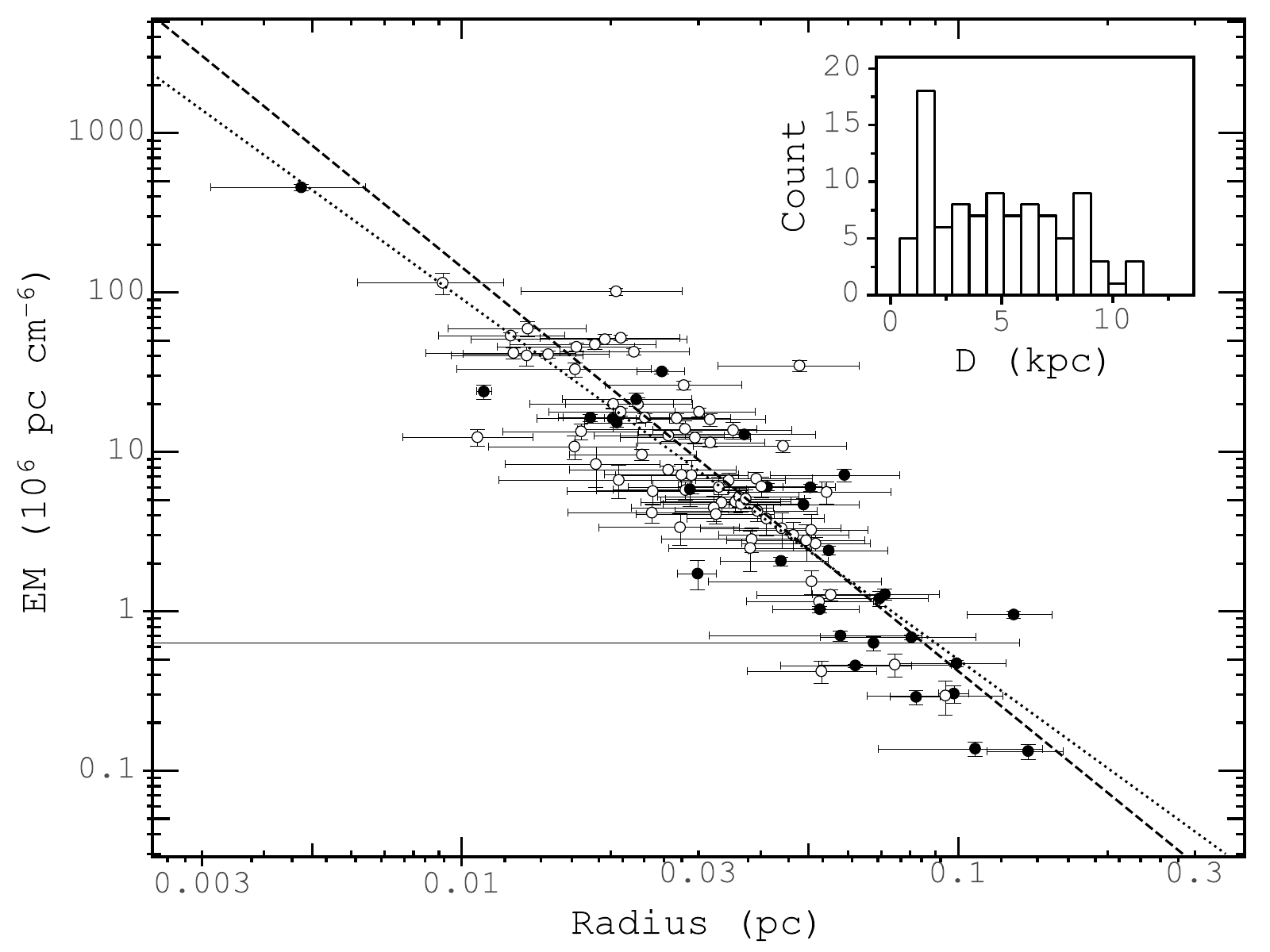}
 \caption{{\bf Main plot:} EM-R distribution of PNe in our sample with known distances. EM and R are from the \spshell\ model. Dashed and dotted lines are the weighted and unweighted best fits respectively. Filled circles represent PNe with independent (calibration) distances and open circles with SB-R distances from FP16. {\bf Subplot:}, in the upper right corner, shows the distribution of distances used for R estimates.}
 \label{fig:emradius}
\end{figure}

As a ``byproduct'' of angular diameter estimates we derived the emission measure EM through the centre of each nebula. The EM, though typically used in UV and X-ray astronomy, can also be used in cases of optically thin radio emission. It is simply defined as the square of the number density of free electrons integrated over the volume of the PN (ie the ionised plasma). Since both values have been simultaneously extracted from the data and the SED fit they are not independent. Confirming that the angular diameter values extracted are accurate when compared to reliable comparison data implies that the EM estimated from the same data might also be accurate (within approximations of the assumed model). 

In Figure~\ref{fig:emradius} we show a plot of EM estimates from the \spshell\ model against measurements of physical PN radii (R) in pc for PNe with decent distance estimates. For calculation of physical diameters we used distance estimates from FP16. Since the Sb-R relation is based on analogous principles to the EM-R relation we separately show a set of independent distance estimates (used as calibrators for the Sb-R relationship) and distance estimates from the Sb-R relationship itself, as filled and open circles, respectively.

The Pearson correlation coefficient of $r=-0.87$ between $\log{\textrm{EM}}$ and $\log{\textrm{R}}$ suggest strong correlation between these two parameters. Since the EM-R relation is essentially the $\textrm{EM}-\theta_{\mathrm{erd}}$ relation convolved with independent and fairly random distribution function of distances this strong correlation is compelling evidence that the EM-R relation is real and it does not stem from the method itself.

The change of EM with the nebular expansion, as evident from this plot, is as expected. By its definition the EM describes the degree of PN expansion that has occurred since the last envelope ejection that created the PN and assuming no additional ionised mass has been contributed from the CSPN. Except for the very early ionisation phase the density of the ionised gas in a PN will gradually decrease as the initially compressed shell expands into the surrounding ISM or indeed into the weakly ionised or neutral wind material that may be present in an extended AGB halo. The power-law fit implies an EM-R relationship of $\textrm{EM}\propto \textrm{R}^{-2.5}$. This result suggests a much shallower relationship between the mean density of the ionised gas (N$_e$to the radius of the nebula ($\textrm{N}_e\propto \textrm{R}^{-1.75}$) than we would expect from a simple expansion with constant nebular mass ($\textrm{N}_e\propto \textrm{R}^{-3}$). However, it is important to note that in this case the estimated slope \underline{is} dependent on the model and approximations used.

Finally, another important value of the correlation found is in using it for distance determination for unresolved objects. All Shklovsky based distance methods (e.g. \cite{Cahn1992, Stanghellini2008, Frew2016}) rely on accurately determined angular diameters which heavily affect its application to all unresolved PNe. In future papers we will examine this possible utilisation of our method by examining its results on sub-samples of PNe with known distances (i.e. those in the Galactic Bulge and Magellanic Clouds or from GAIA determinations of distance for their CSPN).

\section{Conclusions}

An investigation of radio multi-frequency SED fitting as a mechanism for determining reliable angular diameters of PNe is presented. Two simplified models of expected PN spatial emission distribution have been used for this method. This method has a large advantage over conventional methods by not requiring high resolution observations to be used as long as the nearby, unrelated emission is resolved out.

In summary, in comparison with conventional methods we found:

\begin{itemize}
    \item The \spshell\ model results show acceptable agreement with published estimates from other methods. Our values show a small, overall bias of $\approx15$\% that can be accounted for by the approximations used. The spread of Mean Difference values around the mean is comparable to what we could see in comparison between published optical and radio methods. Therefore, we conclude that the precision of our method is similar to those of previous methods. 
    \item The \plshell\ model shows significantly larger bias and it can not be used for angular diameter estimates. This result implies that the ionised, low-density AGB wind does not contribute significantly in the collected overall radio emission from a PN. The reasons behind this result could be twofold. Firstly, from the observational perspective, the fainter and spatially larger emission could be filtered out by insufficient ${\it uv}$ coverage. Secondly, the density distribution in the ionised AGB halo could be much steeper than anticipated from using a simple constant velocity wind approximation.
\end{itemize}

Using our SED method and \spshell\ approximation we provide angular diameters for \NoSEDfittedspshell\ PNe. Also, we measure and provide angular diameters using the 10\% method for \NoNewHSTDiams\ and
\NoNewVLADiams\ GPNe from HST and VLA images, respectively.

In this work we provide a precursor to a more general application to large samples of both Galactic and Local Group PNe for which large-scale radio surveys, covering a wide range of frequencies, such as with the MWA, LOFAR and ASKAP \citep{beardsley2019,shimwell2019,Joseph2019}, are particularly well suited. In addition to improving the coverage of the available radio data for known PNe, these future surveys will certainly contain a significant number of not-yet-detected PNe hidden in their data sets. While no decisive method for distinguishing between PNe and their mimics (compact HII regions, symbiotic systems, etc) based solely on the radio properties is feasible, using machine learning approach on these new data sets in combination with available infrared data (e.g. \citealt{akras2019}), could become a new avenue for improving completeness of Galactic and extragalactic PN populations.

The future power of our newly proposed method will lie in providing robust angular size estimates for distant, compact PNe where ground based optical imaging has insufficient resolution and/or when the optical signal has been too severely attenuated due to dust. 

\section*{Acknowledgements}
The authors would like to thank anonymous referee for constructive comments and suggestions. QAP thanks the Hong Kong Research Grants Council for General Research Fund (GRF) support under grants 17346116 and 17400417. We appreciate Dr Branislav Vukoti\'c for helpful discussion regarding fitting techniques. The NVAS images presented in this paper were produced as part of the NRAO VLA Archive Survey, (c) AUI/NRAO. This research made use of Astropy,\footnote{http://www.astropy.org} a community-developed core Python package for Astronomy \citep{robitaille2013, theastropycollaboration2018}. This research was made based on observations made with the NASA/ESA Hubble Space Telescope, and obtained from the Hubble Legacy Archive, which is a collaboration between the Space Telescope Science Institute (STScI/NASA), the Space Telescope European Coordinating Facility (ST-ECF/ESA) and the Canadian Astronomy Data Centre (CADC/NRC/CSA). The International Centre for Radio Astronomy Research (ICRAR) is a Joint Venture of Curtin University and The University of Western Australia, funded by the Western Australian State government. This research made use of APLpy, an open-source plotting package for Python \citep{robitaille2012}. This research has made use of the TOPCAT, an interactive graphical viewer and editor for tabular data \citep{topcat} and HASH PN database at \href{http://hashpn.space}{hashpn.space}.

\section{Data Availability}

The data underlying this article are available through public archives and in the article and in its online supplementary material.

%%%%%%%%%%%%%%%%%%%%%%%%%%%%%%%%%%%%%%%%%%%%%%%%%%



\bibliographystyle{mnras}
\bibliography{SED_PNe.bib}

\begin{thebibliography}{}
\makeatletter
\relax
\def\mn@urlcharsother{\let\do\@makeother \do\$\do\&\do\#\do\^\do\_\do\%\do\~}
\def\mn@doi{\begingroup\mn@urlcharsother \@ifnextchar [ {\mn@doi@}
  {\mn@doi@[]}}
\def\mn@doi@[#1]#2{\def\@tempa{#1}\ifx\@tempa\@empty \href
  {http://dx.doi.org/#2} {doi:#2}\else \href {http://dx.doi.org/#2} {#1}\fi
  \endgroup}
\def\mn@eprint#1#2{\mn@eprint@#1:#2::\@nil}
\def\mn@eprint@arXiv#1{\href {http://arxiv.org/abs/#1} {{\tt arXiv:#1}}}
\def\mn@eprint@dblp#1{\href {http://dblp.uni-trier.de/rec/bibtex/#1.xml}
  {dblp:#1}}
\def\mn@eprint@#1:#2:#3:#4\@nil{\def\@tempa {#1}\def\@tempb {#2}\def\@tempc
  {#3}\ifx \@tempc \@empty \let \@tempc \@tempb \let \@tempb \@tempa \fi \ifx
  \@tempb \@empty \def\@tempb {arXiv}\fi \@ifundefined
  {mn@eprint@\@tempb}{\@tempb:\@tempc}{\expandafter \expandafter \csname
  mn@eprint@\@tempb\endcsname \expandafter{\@tempc}}}

\bibitem[\protect\citeauthoryear{Aaquist \& Kwok}{Aaquist \&
  Kwok}{1990}]{Aaquist1990}
Aaquist O.~B.,  Kwok S.,  1990, A\&AS, \href
  {http://adsabs.harvard.edu/abs/1990A\%26AS...84..229A} {84, 229}

\bibitem[\protect\citeauthoryear{Aaquist \& Kwok}{Aaquist \&
  Kwok}{1991}]{Aaquist1991}
Aaquist O.~B.,  Kwok S.,  1991, \mn@doi [ApJ] {10.1086/170461}, \href
  {http://adsabs.harvard.edu/abs/1991ApJ...378..599A} {378, 599}

\bibitem[\protect\citeauthoryear{Acker, Marcout, Ochsenbein, Stenholm, Tylenda
  \& Schohn}{Acker et~al.}{1992}]{Acker1992a}
Acker A.,  Marcout J.,  Ochsenbein F.,  Stenholm B.,  Tylenda R.,   Schohn C.,
  1992, The Strasbourg-ESO Catalogue of Galactic Planetary Nebulae. Parts I,
  II., by Acker, A.; Marcout, J.; Ochsenbein, F.; Stenholm, B.; Tylenda, R.;
  Schohn, C.. European Southern Observatory, Garching (Germany), 1992, 1047 p.,
  ISBN 3-923524-41-2,

\bibitem[\protect\citeauthoryear{{Akras}, {Guzman-Ramirez}  \&
  {Gon{\c{c}}alves}}{{Akras} et~al.}{2019}]{akras2019}
{Akras} S.,  {Guzman-Ramirez} L.,   {Gon{\c{c}}alves} D.~R.,  2019, \mn@doi
  [\mnras] {10.1093/mnras/stz1911}, \href
  {https://ui.adsabs.harvard.edu/abs/2019MNRAS.488.3238A} {488, 3238}

\bibitem[\protect\citeauthoryear{Aller \& Milne}{Aller \&
  Milne}{1972}]{Aller1972a}
Aller L.~H.,  Milne D.~K.,  1972, Australian Journal of Physics, \href
  {http://adsabs.harvard.edu/abs/1972AuJPh..25...91A} {25, 91}

\bibitem[\protect\citeauthoryear{Anderson, Zavagno, Barlow, {Garc{\'i}a-Lario}
  \& {Noriega-Crespo}}{Anderson et~al.}{2012}]{Anderson2012}
Anderson L.~D.,  Zavagno A.,  Barlow M.~J.,  {Garc{\'i}a-Lario} P.,
  {Noriega-Crespo} A.,  2012, \mn@doi [A\&A] {10.1051/0004-6361/201117640},
  537, A1

\bibitem[\protect\citeauthoryear{Bains, Bryce, Mellema, Redman  \&
  Thomasson}{Bains et~al.}{2003}]{Bains2003}
Bains I.,  Bryce M.,  Mellema G.,  Redman M.~P.,   Thomasson P.,  2003, \mn@doi
  [MNRAS] {10.1046/j.1365-8711.2003.06216.x}, \href
  {http://adsabs.harvard.edu/abs/2003MNRAS.340..381B} {340, 381}

\bibitem[\protect\citeauthoryear{Basart \& Daub}{Basart \&
  Daub}{1987}]{Basart1987}
Basart J.~P.,  Daub C.~T.,  1987, \mn@doi [ApJ] {10.1086/165286}, \href
  {http://adsabs.harvard.edu/abs/1987ApJ...317..412B} {317, 412}

\bibitem[\protect\citeauthoryear{Beardsley et~al.,}{Beardsley
  et~al.}{2019}]{beardsley2019}
Beardsley A.~P.,  et~al., 2019, \mn@doi [Publications of the Astronomical
  Society of Australia] {10.1017/pasa.2019.41}, 36, e050

\bibitem[\protect\citeauthoryear{Becker, White  \& Edwards}{Becker
  et~al.}{1991}]{Becker1991}
Becker R.~H.,  White R.~L.,   Edwards A.~L.,  1991, \mn@doi [A\&AS]
  {10.1086/191529}, 75, 1

\bibitem[\protect\citeauthoryear{Bedding \& Zijlstra}{Bedding \&
  Zijlstra}{1994}]{Bedding1994}
Bedding T.~R.,  Zijlstra A.~A.,  1994, A\&A, 283, 955

\bibitem[\protect\citeauthoryear{Bensby \& Lundstr{\"o}m}{Bensby \&
  Lundstr{\"o}m}{2001}]{Bensby2001}
Bensby T.,  Lundstr{\"o}m I.,  2001, \mn@doi [A\&A]
  {10.1051/0004-6361:20010705}, \href
  {http://adsabs.harvard.edu/abs/2001A\%26A...374..599B} {374, 599}

\bibitem[\protect\citeauthoryear{Bland \& Altman}{Bland \&
  Altman}{1999}]{bland1999}
Bland J.~M.,  Altman D.~G.,  1999, \mn@doi [Statistical Methods in Medical
  Research; London]
  {http://dx.doi.org.ezproxy.uws.edu.au/10.1191/096228099673819272}, 8, 135

\bibitem[\protect\citeauthoryear{Bojicic}{Bojicic}{2010}]{Bojicic2010a}
Bojicic I.,  2010, PhD thesis, Macquarie University

\bibitem[\protect\citeauthoryear{Boji{\v c}i{\'c}, Parker, Filipovi{\'c}  \&
  Frew}{Boji{\v c}i{\'c} et~al.}{2011}]{bojicic2011}
Boji{\v c}i{\'c} I.~S.,  Parker Q.~A.,  Filipovi{\'c} M.~D.,   Frew D.~J.,
  2011, \mn@doi [MNRAS] {10.1111/j.1365-2966.2010.17900.x}, 412, 223

\bibitem[\protect\citeauthoryear{Bojicic, Parker  \& Frew}{Bojicic
  et~al.}{2017}]{Bojicic2017}
Bojicic I.~S.,  Parker Q.~A.,   Frew D.~J.,  2017, in Liu X.,  Stanghellini L.,
    Karakas A.,  eds,  {{IAU}} Symposium Vol. 323, Planetary Nebulae:
  {{Multi}}-Wavelength Probes of Stellar and Galactic Evolution. pp 327--328,
  \mn@doi{10.1017/S1743921317003234}

\bibitem[\protect\citeauthoryear{Branch, Coleman  \& Li}{Branch
  et~al.}{1999}]{branch1999}
Branch M.~A.,  Coleman T.~F.,   Li Y.,  1999, \mn@doi [SIAM Journal on
  Scientific Computing] {10.1137/S1064827595289108}, 21, 1

\bibitem[\protect\citeauthoryear{Cahn, Kaler  \& Stanghellini}{Cahn
  et~al.}{1992}]{Cahn1992}
Cahn J.~H.,  Kaler J.~B.,   Stanghellini L.,  1992, A\&AS, 94, 399

\bibitem[\protect\citeauthoryear{Calabretta}{Calabretta}{1982}]{Calabretta1982}
Calabretta M.~R.,  1982, MNRAS, \href
  {http://adsabs.harvard.edu/abs/1982MNRAS.199..141C} {199, 141}

\bibitem[\protect\citeauthoryear{Cerrigone, Hora, Umana  \& Trigilio}{Cerrigone
  et~al.}{2008}]{Cerrigone2008}
Cerrigone L.,  Hora J.~L.,  Umana G.,   Trigilio C.,  2008, \mn@doi [ApJ]
  {10.1086/589228}, \href {http://ads.nao.ac.jp/abs/2008ApJ...682.1047C} {682,
  1047}

\bibitem[\protect\citeauthoryear{Collaboration et~al.,}{Collaboration
  et~al.}{2018}]{theastropycollaboration2018}
Collaboration T.~A.,  et~al., 2018, \mn@doi [AJ] {10.3847/1538-3881/aabc4f},
  156, 123

\bibitem[\protect\citeauthoryear{Condon \& Kaplan}{Condon \&
  Kaplan}{1998}]{Condon1998a}
Condon J.~J.,  Kaplan D.~L.,  1998, \mn@doi [A\&AS] {10.1086/313128}, 117, 361

\bibitem[\protect\citeauthoryear{Condon, Cotton, Greisen, Yin, Perley, Taylor
  \& Broderick}{Condon et~al.}{1998}]{Condon1998}
Condon J.~J.,  Cotton W.~D.,  Greisen E.~W.,  Yin Q.~F.,  Perley R.~A.,  Taylor
  G.~B.,   Broderick J.~J.,  1998, \mn@doi [AJ] {10.1086/300337}, \href
  {http://adsabs.harvard.edu/abs/1998AJ....115.1693C} {115, 1693}

\bibitem[\protect\citeauthoryear{Cornwell}{Cornwell}{1988}]{Cornwell1988}
Cornwell T.~J.,  1988, A\&A, \href
  {http://adsabs.harvard.edu/abs/1988A\%26A...202..316C} {202, 316}

\bibitem[\protect\citeauthoryear{De~Breuck, Tang, {de Bruyn}, R{\"o}ttgering
  \& {van Breugel}}{De~Breuck et~al.}{2002}]{DeBreuck2002}
De~Breuck C.,  Tang Y.,  {de Bruyn} A.~G.,  R{\"o}ttgering H.,   {van Breugel}
  W.,  2002, \mn@doi [A\&A] {10.1051/0004-6361:20021115}, 394, 59

\bibitem[\protect\citeauthoryear{Douglas, Bash, Bozyan, Torrence  \&
  Wolfe}{Douglas et~al.}{1996}]{Douglas1996}
Douglas J.~N.,  Bash F.~N.,  Bozyan F.~A.,  Torrence G.~W.,   Wolfe C.,  1996,
  \mn@doi [AJ] {10.1086/117932}, 111, 1945

\bibitem[\protect\citeauthoryear{Fragkou, Parker, Boji{\v c}i{\'c}  \&
  Aksaker}{Fragkou et~al.}{2018}]{Fragkou2018}
Fragkou V.,  Parker Q.~A.,  Boji{\v c}i{\'c} I.~S.,   Aksaker N.,  2018,
  \mn@doi [MNRAS] {10.1093/mnras/sty1977}, 480, 2916

\bibitem[\protect\citeauthoryear{Frew \& Parker}{Frew \&
  Parker}{2010}]{Frew2010a}
Frew D.~J.,  Parker Q.~A.,  2010, \mn@doi [PASA] {10.1071/AS09040}, \href
  {http://ads.nao.ac.jp/abs/2010PASA...27..129F} {27, 129}

\bibitem[\protect\citeauthoryear{Frew, Parker  \& Boji{\v c}i{\'c}}{Frew
  et~al.}{2016}]{Frew2016}
Frew D.~J.,  Parker Q.~A.,   Boji{\v c}i{\'c} I.~S.,  2016, \mn@doi [MNRAS]
  {10.1093/mnras/stv1516}, 455, 1459

\bibitem[\protect\citeauthoryear{Furst, Reich, Reich  \& Reif}{Furst
  et~al.}{1990}]{Furst1990}
Furst E.,  Reich W.,  Reich P.,   Reif K.,  1990, A\&AS, 85, 805

\bibitem[\protect\citeauthoryear{Garwood, Perley, Dickey  \& Murray}{Garwood
  et~al.}{1988}]{Garwood1998}
Garwood R.~W.,  Perley R.~A.,  Dickey J.~M.,   Murray M.~A.,  1988, \mn@doi
  [AJ] {10.1086/114917}, 96, 1655

\bibitem[\protect\citeauthoryear{Gathier, Pottasch, Goss  \& {van
  Gorkom}}{Gathier et~al.}{1983}]{Gathier1983}
Gathier R.,  Pottasch S.~R.,  Goss W.~M.,   {van Gorkom} J.~H.,  1983, A\&A,
  128, 325

\bibitem[\protect\citeauthoryear{Gledhill, Froebrich, {Campbell-White}  \&
  Jones}{Gledhill et~al.}{2018}]{Gledhill2018}
Gledhill T.~M.,  Froebrich D.,  {Campbell-White} J.,   Jones A.~M.,  2018,
  \mn@doi [MNRAS] {10.1093/mnras/sty1580}, 479, 3759

\bibitem[\protect\citeauthoryear{Gorny, Schwarz, Corradi  \& Van~Winckel}{Gorny
  et~al.}{1999}]{Gorny1999}
Gorny S.~K.,  Schwarz H.~E.,  Corradi R.~L.,   Van~Winckel H.,  1999, \mn@doi
  [A\&AS] {10.1051/aas:1999205}, 136, 145

\bibitem[\protect\citeauthoryear{Grubbs}{Grubbs}{1969}]{Grubbs1969}
Grubbs F.~E.,  1969, Technometrics, 11, 1

\bibitem[\protect\citeauthoryear{Hajduk, {van Hoof}, {\'S}niadkowska,
  Krankowski, B{\l}aszkiewicz, D{\k{a}}browski  \& Zijlstra}{Hajduk
  et~al.}{2018}]{Hajduk2018}
Hajduk M.,  {van Hoof} P. A.~M.,  {\'S}niadkowska K.,  Krankowski A.,
  B{\l}aszkiewicz L.,  D{\k{a}}browski B.,   Zijlstra A.~A.,  2018, \mn@doi
  [Monthly Notices of the Royal Astronomical Society] {10.1093/mnras/sty1673},
  479, 5657

\bibitem[\protect\citeauthoryear{{Harvey-Smith}, Hardwick, De~Marco,
  Parthasarathy, Gonidakis, Akhter, Cunningham  \& Green}{{Harvey-Smith}
  et~al.}{2018}]{Harvey-Smith2018}
{Harvey-Smith} L.,  Hardwick J.~A.,  De~Marco O.,  Parthasarathy M.,  Gonidakis
  I.,  Akhter S.,  Cunningham M.,   Green J.~A.,  2018, \mn@doi [MNRAS]
  {10.1093/mnras/sty1513}, 479, 1842

\bibitem[\protect\citeauthoryear{Helfand, Becker, White, Fallon  \&
  Tuttle}{Helfand et~al.}{2006}]{Helfand2006}
Helfand D.~J.,  Becker R.~H.,  White R.~L.,  Fallon A.,   Tuttle S.,  2006,
  \mn@doi [AJ] {10.1086/503253}, \href
  {http://ads.nao.ac.jp/abs/2006AJ....131.2525H} {131, 2525}

\bibitem[\protect\citeauthoryear{Hoare et~al.,}{Hoare et~al.}{2012}]{Hoare2012}
Hoare M.~G.,  et~al., 2012, \mn@doi [PASP] {10.1086/668058}, 124, 939,\"A\`i955

\bibitem[\protect\citeauthoryear{{Hurley-Walker} et~al.,}{{Hurley-Walker}
  et~al.}{2017}]{hurleywalker2017}
{Hurley-Walker} N.,  et~al., 2017, \mn@doi [MNRAS] {10.1093/mnras/stw2337},
  464, 1146

\bibitem[\protect\citeauthoryear{{Hurley-Walker} et~al.,}{{Hurley-Walker}
  et~al.}{2019}]{hurleywalker2019}
{Hurley-Walker} N.,  et~al., 2019, \mn@doi [PASA] {10.1017/pasa.2019.37}, 36,
  e047

\bibitem[\protect\citeauthoryear{Intema, Jagannathan, Mooley  \& Frail}{Intema
  et~al.}{2017}]{Intema2017}
Intema H.~T.,  Jagannathan P.,  Mooley K.~P.,   Frail D.~A.,  2017, \mn@doi
  [A\&A] {10.1051/0004-6361/201628536}, 598, A78

\bibitem[\protect\citeauthoryear{Irabor et~al.,}{Irabor
  et~al.}{2018}]{Irabor2018}
Irabor T.,  et~al., 2018, \mn@doi [MNRAS] {10.1093/mnras/sty1929}, 480, 2423

\bibitem[\protect\citeauthoryear{Isaacman}{Isaacman}{1984}]{Isaacman1984}
Isaacman R.,  1984, MNRAS, 208, 399

\bibitem[\protect\citeauthoryear{Jacoby}{Jacoby}{1980}]{Jacoby1980}
Jacoby G.~H.,  1980, \mn@doi [A\&AS] {10.1086/190642}, \href
  {http://adsabs.harvard.edu/abs/1980ApJS...42....1J} {42, 1}

\bibitem[\protect\citeauthoryear{{Joseph} et~al.,}{{Joseph}
  et~al.}{2019}]{Joseph2019}
{Joseph} T.~D.,  et~al., 2019, \mn@doi [\mnras] {10.1093/mnras/stz2650}, \href
  {https://ui.adsabs.harvard.edu/abs/2019MNRAS.490.1202J} {490, 1202}

\bibitem[\protect\citeauthoryear{{Kimeswenger} \& {Barr{\'\i}a}}{{Kimeswenger}
  \& {Barr{\'\i}a}}{2018}]{Kimeswenger2018}
{Kimeswenger} S.,  {Barr{\'\i}a} D.,  2018, \mn@doi [\aap]
  {10.1051/0004-6361/201833647}, \href
  {https://ui.adsabs.harvard.edu/abs/2018A&A...616L...2K} {616, L2}

\bibitem[\protect\citeauthoryear{Kovacevic, Parker, Jacoby, Sharp, Miszalski
  \& Frew}{Kovacevic et~al.}{2011}]{Kovacevic2011}
Kovacevic A.~V.,  Parker Q.~A.,  Jacoby G.~H.,  Sharp R.,  Miszalski B.,   Frew
  D.~J.,  2011, \mn@doi [MNRAS] {10.1111/j.1365-2966.2011.18250.x}, \href
  {http://ads.nao.ac.jp/abs/2011MNRAS.414..860K} {414, 860}

\bibitem[\protect\citeauthoryear{Kronberger et~al.,}{Kronberger
  et~al.}{2016}]{Kronberger2016}
Kronberger M.,  et~al., 2016, \mn@doi [Journal of Physics Conference Series]
  {10.1088/1742-6596/728/7/072012}, \href
  {http://adsabs.harvard.edu/abs/2016JPhCS.728g2012K} {728, 072012}

\bibitem[\protect\citeauthoryear{Kwok}{Kwok}{1985}]{Kwok1985a}
Kwok S.,  1985, \mn@doi [AJ] {10.1086/113707}, 90, 49

\bibitem[\protect\citeauthoryear{Kwok \& Aaquist}{Kwok \&
  Aaquist}{1993}]{Kwok1993a}
Kwok S.,  Aaquist O.~B.,  1993, PASP, \href
  {http://adsabs.harvard.edu/abs/1993PASP..105.1456K} {105, 1456}

\bibitem[\protect\citeauthoryear{Kwok, Purton  \& Keenan}{Kwok
  et~al.}{1981}]{Kwok1981}
Kwok S.,  Purton C.~R.,   Keenan D.~W.,  1981, \mn@doi [ApJ] {10.1086/159367},
  250, 232

\bibitem[\protect\citeauthoryear{Lynds}{Lynds}{1963}]{Lynds1963}
Lynds C.~R.,  1963, National Radio Astronomy Observatory Publications, 1, 85

\bibitem[\protect\citeauthoryear{Marigo, Girardi, Groenewegen  \& Weiss}{Marigo
  et~al.}{2001}]{Marigo2001}
Marigo P.,  Girardi L.,  Groenewegen M. A.~T.,   Weiss A.,  2001, \mn@doi
  [A\&A] {10.1051/0004-6361:20011270}, \href
  {http://adsabs.harvard.edu/abs/2001A\%26A...378..958M} {378, 958}

\bibitem[\protect\citeauthoryear{Marten \& Schoenberner}{Marten \&
  Schoenberner}{1991}]{Marten1991}
Marten H.,  Schoenberner D.,  1991, A\&A, \href
  {http://adsabs.harvard.edu/abs/1991A\%26A...248..590M} {248, 590}

\bibitem[\protect\citeauthoryear{Mauch, Murphy, Buttery, Curran, Hunstead,
  Piestrzynski, Robertson  \& Sadler}{Mauch et~al.}{2003}]{Mauch2003}
Mauch T.,  Murphy T.,  Buttery H.~J.,  Curran J.,  Hunstead R.~W.,
  Piestrzynski B.,  Robertson J.~G.,   Sadler E.~M.,  2003, \mn@doi [MNRAS]
  {10.1046/j.1365-8711.2003.06605.x}, \href
  {http://cdsads.u-strasbg.fr/abs/2003MNRAS.342.1117M} {342, 1117}

\bibitem[\protect\citeauthoryear{McConnell, Sadler, Murphy  \& Ekers}{McConnell
  et~al.}{2012}]{McConnell2012}
McConnell D.,  Sadler E.~M.,  Murphy T.,   Ekers R.~D.,  2012, \mn@doi [MNRAS]
  {10.1111/j.1365-2966.2012.20726.x}, 422, 1527

\bibitem[\protect\citeauthoryear{Mellema}{Mellema}{1994}]{Mellema1994}
Mellema G.,  1994, A\&A, \href
  {http://cdsads.u-strasbg.fr/abs/1994A\%26A...290..915M} {290, 915}

\bibitem[\protect\citeauthoryear{Menon \& Terzian}{Menon \&
  Terzian}{1965}]{Menon1965}
Menon T.~K.,  Terzian Y.,  1965, \mn@doi [ApJ] {10.1086/148158}, \href
  {http://adsabs.harvard.edu/abs/1965ApJ...141..745M} {141, 745}

\bibitem[\protect\citeauthoryear{Meyers et~al.,}{Meyers
  et~al.}{2017}]{Meyers2017}
Meyers B.~W.,  et~al., 2017, \mn@doi [PASA] {10.1017/pasa.2017.5}, 34, e013

\bibitem[\protect\citeauthoryear{Milne}{Milne}{1979}]{Milne1979}
Milne D.~K.,  1979, A\&AS, \href
  {http://adsabs.harvard.edu/abs/1979A\%26AS...36..227M} {36, 227}

\bibitem[\protect\citeauthoryear{Milne \& Aller}{Milne \&
  Aller}{1975}]{Milne1975}
Milne D.~K.,  Aller L.~H.,  1975, A\&A, 38, 183

\bibitem[\protect\citeauthoryear{Milne \& Aller}{Milne \&
  Aller}{1982}]{Milne1982}
Milne D.~K.,  Aller L.~H.,  1982, A\&AS, \href
  {http://adsabs.harvard.edu/abs/1982A\%26AS...50..209M} {50, 209}

\bibitem[\protect\citeauthoryear{Moe \& De~Marco}{Moe \&
  De~Marco}{2006}]{Moe2006}
Moe M.,  De~Marco O.,  2006, \mn@doi [ApJ] {10.1086/506900}, 650, 916

\bibitem[\protect\citeauthoryear{Mross, Weinberger  \& Hartl}{Mross
  et~al.}{1981}]{Mross1981}
Mross R.,  Weinberger R.,   Hartl H.,  1981, A\&AS, \href
  {http://adsabs.harvard.edu/abs/1981A\%26AS...43...75M} {43, 75}

\bibitem[\protect\citeauthoryear{Murphy, Mauch, Green, {Hunstead, R. W.
  andPiestrzynska}, Kels  \& Sztajer}{Murphy et~al.}{2007}]{Murphy2007}
Murphy T.,  Mauch T.,  Green A.,  {Hunstead, R. W. andPiestrzynska} B.,  Kels
  A.~P.,   Sztajer P.,  2007, \mn@doi [MNRAS]
  {10.1111/j.1365-2966.2007.12379.x}, \href
  {http://cdsads.u-strasbg.fr/abs/2007MNRAS.382..382M} {382, 382}

\bibitem[\protect\citeauthoryear{Murphy et~al.,}{Murphy
  et~al.}{2010}]{Murphy2010a}
Murphy T.,  et~al., 2010, \mn@doi [MNRAS] {10.1111/j.1365-2966.2009.15961.x},
  \href {http://ads.nao.ac.jp/abs/2010MNRAS.402.2403M} {402, 2403}

\bibitem[\protect\citeauthoryear{Nord, Lazio, Kassim, Hyman, LaRosa, Brogan  \&
  Duric}{Nord et~al.}{2004}]{Nord2004}
Nord M.~E.,  Lazio T. J.~W.,  Kassim N.~E.,  Hyman S.~D.,  LaRosa T.~N.,
  Brogan C.~L.,   Duric N.,  2004, \mn@doi [AJ] {10.1086/424001}, 128, 1646

\bibitem[\protect\citeauthoryear{Olnon}{Olnon}{1975}]{Olnon1975}
Olnon F.~M.,  1975, A\&A, 39, 217

\bibitem[\protect\citeauthoryear{Parker et~al.,}{Parker
  et~al.}{2012}]{Parker2012}
Parker Q.~A.,  et~al., 2012, \mn@doi [MNRAS]
  {10.1111/j.1365-2966.2012.21927.x}, 427, 3016

\bibitem[\protect\citeauthoryear{Parker, Bojicic  \& Frew}{Parker
  et~al.}{2016}]{Parker2016}
Parker Q.~A.,  Bojicic I.,   Frew D.~J.,  2016, preprint, \href
  {http://adsabs.harvard.edu/abs/2016arXiv160307042P} {}

\bibitem[\protect\citeauthoryear{Pazderska et~al.,}{Pazderska
  et~al.}{2009}]{Pazderska2009}
Pazderska B.~M.,  et~al., 2009, \mn@doi [A\&A] {10.1051/0004-6361/200811369},
  \href {http://adsabs.harvard.edu/abs/2009A\%26A...498..463P} {498, 463}

\bibitem[\protect\citeauthoryear{Perinotto, Sch{\"o}nberner, Steffen  \&
  Calonaci}{Perinotto et~al.}{2004}]{Perinotto2004}
Perinotto M.,  Sch{\"o}nberner D.,  Steffen M.,   Calonaci C.,  2004, \mn@doi
  [A\&A] {10.1051/0004-6361:20031653}, \href
  {http://adsabs.harvard.edu/abs/2004A\%26A...414..993P} {414, 993}

\bibitem[\protect\citeauthoryear{Pety \& {Rodr{\'i}guez-Fern{\'a}ndez}}{Pety \&
  {Rodr{\'i}guez-Fern{\'a}ndez}}{2010}]{pety2010}
Pety J.,  {Rodr{\'i}guez-Fern{\'a}ndez} N.,  2010, \mn@doi [A\&A]
  {10.1051/0004-6361/200912873}, 517, A12

\bibitem[\protect\citeauthoryear{Phillips \& Mampaso}{Phillips \&
  Mampaso}{1988}]{Phillips1988}
Phillips J.~P.,  Mampaso A.,  1988, A\&A, \href
  {http://adsabs.harvard.edu/abs/1988A\%26A...190..237P} {190, 237}

\bibitem[\protect\citeauthoryear{Pottasch}{Pottasch}{1984}]{Pottasch1984}
Pottasch S.~R.,  1984, in Astrophysics and Space Science Library, Vol.~107,
  Astrophysics and Space Science Library.
{D. Reidel}

\bibitem[\protect\citeauthoryear{Pottasch \& Zijlstra}{Pottasch \&
  Zijlstra}{1992}]{Pottasch1992}
Pottasch S.~R.,  Zijlstra A.~A.,  1992, A\&A, 256, 251

\bibitem[\protect\citeauthoryear{Pottasch \& Zijlstra}{Pottasch \&
  Zijlstra}{1994}]{Pottasch1994}
Pottasch S.~R.,  Zijlstra A.~A.,  1994, A\&A, \href
  {http://adsabs.harvard.edu/abs/1994A\%26A...289..261P} {289, 261}

\bibitem[\protect\citeauthoryear{Purton, Feldman, Marsh, Allen  \&
  Wright}{Purton et~al.}{1982}]{Purton1982}
Purton C.~R.,  Feldman P.~A.,  Marsh K.~A.,  Allen D.~A.,   Wright A.~E.,
  1982, MNRAS, \href {http://adsabs.harvard.edu/abs/1982MNRAS.198..321P} {198,
  321}

\bibitem[\protect\citeauthoryear{Ratag \& Pottasch}{Ratag \&
  Pottasch}{1991}]{Ratag1991}
Ratag M.~A.,  Pottasch S.~R.,  1991, A\&AS, \href
  {http://adsabs.harvard.edu/abs/1991A\%26AS...91..481R} {91, 481}

\bibitem[\protect\citeauthoryear{Ratag, Pottasch, Zijlstra  \& Menzies}{Ratag
  et~al.}{1990}]{Ratag1990}
Ratag M.~A.,  Pottasch S.~R.,  Zijlstra A.~A.,   Menzies J.,  1990, A\&A, \href
  {http://adsabs.harvard.edu/abs/1990A\%26A...233..181R} {233, 181}

\bibitem[\protect\citeauthoryear{Reich, Fuerst, Haslam, Steffen  \& Reif}{Reich
  et~al.}{1984}]{Reich1984}
Reich W.,  Fuerst E.,  Haslam C. G.~T.,  Steffen P.,   Reif K.,  1984, A\&AS,
  \href {http://cdsads.u-strasbg.fr/abs/1984A\%26AS...58..197R} {58, 197}

\bibitem[\protect\citeauthoryear{Rengelink, Tang, {de Bruyn}, Miley, Bremer,
  Roettgering  \& Bremer}{Rengelink et~al.}{1997}]{Rengelink1997}
Rengelink R.~B.,  Tang Y.,  {de Bruyn} A.~G.,  Miley G.~K.,  Bremer M.~N.,
  Roettgering H. J.~A.,   Bremer M. A.~R.,  1997, \mn@doi [A\&AS]
  {10.1051/aas:1997358}, 124, 259

\bibitem[\protect\citeauthoryear{{Robitaille} \& {Bressert}}{{Robitaille} \&
  {Bressert}}{2012}]{robitaille2012}
{Robitaille} T.,  {Bressert} E.,  2012, {APLpy: Astronomical Plotting Library
  in Python} (\mn@eprint {ascl} {1208.017})

\bibitem[\protect\citeauthoryear{Robitaille et~al.,}{Robitaille
  et~al.}{2013}]{robitaille2013}
Robitaille T.~P.,  et~al., 2013, \mn@doi [A\&A] {10.1051/0004-6361/201322068},
  558, A33

\bibitem[\protect\citeauthoryear{Rodriguez et~al.,}{Rodriguez
  et~al.}{1985}]{Rodriguez1985}
Rodriguez L.~F.,  et~al., 1985, MNRAS, \href
  {http://adsabs.harvard.edu/abs/1985MNRAS.215..353R} {215, 353}

\bibitem[\protect\citeauthoryear{Ruffle, Zijlstra, Walsh, Gray, Gesicki,
  Minniti  \& Comeron}{Ruffle et~al.}{2004}]{Ruffle2004}
Ruffle P. M.~E.,  Zijlstra A.~A.,  Walsh J.~R.,  Gray M.~D.,  Gesicki K.,
  Minniti D.,   Comeron F.,  2004, \mn@doi [MNRAS]
  {10.1111/j.1365-2966.2004.08113.x}, 353, 796

\bibitem[\protect\citeauthoryear{{Schmidt-Voigt} \& Koeppen}{{Schmidt-Voigt} \&
  Koeppen}{1987}]{Schmidt-Voigt1987}
{Schmidt-Voigt} M.,  Koeppen J.,  1987, A\&A, \href
  {http://adsabs.harvard.edu/abs/1987A\%26A...174..211S} {174, 211}

\bibitem[\protect\citeauthoryear{Schonberner, Jacob, Steffen  \&
  Sandin}{Schonberner et~al.}{2007}]{Schonberner2007}
Schonberner D.,  Jacob R.,  Steffen M.,   Sandin C.,  2007, \mn@doi [A\&A]
  {10.1051/0004-6361:20077437}, \href
  {http://adsabs.harvard.edu/abs/2007A\%26A...473..467S} {473, 467}

\bibitem[\protect\citeauthoryear{Schwarz, Corradi  \& Melnick}{Schwarz
  et~al.}{1992}]{SCHWARZ1992}
Schwarz H.~E.,  Corradi R. L.~M.,   Melnick J.,  1992, A\&AS, 96, 23

\bibitem[\protect\citeauthoryear{Seaquist \& Davis}{Seaquist \&
  Davis}{1983}]{Seaquist1983}
Seaquist E.~R.,  Davis L.~E.,  1983, \mn@doi [ApJ] {10.1086/161478}, \href
  {http://adsabs.harvard.edu/abs/1983ApJ...274..659S} {274, 659}

\bibitem[\protect\citeauthoryear{Shaw, Stanghellini, Villaver  \&
  Mutchler}{Shaw et~al.}{2006}]{Shaw2006}
Shaw R.~A.,  Stanghellini L.,  Villaver E.,   Mutchler M.,  2006, \mn@doi
  [ApJS] {10.1086/508469}, 167, 201,\"A\`i229

\bibitem[\protect\citeauthoryear{Shimwell et~al.,}{Shimwell
  et~al.}{2019}]{shimwell2019}
Shimwell T.~W.,  et~al., 2019, \mn@doi [A\&A] {10.1051/0004-6361/201833559},
  622, A1

\bibitem[\protect\citeauthoryear{Slee \& Orchiston}{Slee \&
  Orchiston}{1965}]{Slee1965}
Slee O.~B.,  Orchiston D.~W.,  1965, Australian Journal of Physics, \href
  {http://adsabs.harvard.edu/abs/1965AuJPh..18..187S} {18, 187}

\bibitem[\protect\citeauthoryear{Stanghellini, Shaw, Balick, Mutchler, Blades
  \& Villaver}{Stanghellini et~al.}{2003}]{Stanghellini2003}
Stanghellini L.,  Shaw R.~A.,  Balick B.,  Mutchler M.,  Blades J.~C.,
  Villaver E.,  2003, \mn@doi [ApJ] {10.1086/378042}, \href
  {http://ads.nao.ac.jp/abs/2003ApJ...596..997S} {596, 997}

\bibitem[\protect\citeauthoryear{Stanghellini, Shaw  \& Villaver}{Stanghellini
  et~al.}{2008}]{Stanghellini2008}
Stanghellini L.,  Shaw R.~A.,   Villaver E.,  2008, \mn@doi [ApJ]
  {10.1086/592395}, \href {http://adsabs.harvard.edu/abs/2008ApJ...689..194S}
  {689, 194}

\bibitem[\protect\citeauthoryear{Stasi{\'n}ska, Tylenda, Acker  \&
  Stenholm}{Stasi{\'n}ska et~al.}{1991}]{Stasinska1991a}
Stasi{\'n}ska G.,  Tylenda R.,  Acker A.,   Stenholm B.,  1991, A\&A, 247, 173

\bibitem[\protect\citeauthoryear{{Stenborg}}{{Stenborg}}{2017}]{stenborg2017}
{Stenborg} T.~N.,  2017, \mn@doi [\pasp] {10.1088/1538-3873/aa6ca4}, \href
  {https://ui.adsabs.harvard.edu/abs/2017PASP..129f7002S} {129, 067002}

\bibitem[\protect\citeauthoryear{{Taylor}}{{Taylor}}{2005}]{topcat}
{Taylor} M.~B.,  2005, in {Shopbell} P.,  {Britton} M.,   {Ebert} R.,  eds,
  Astronomical Society of the Pacific Conference Series Vol. 347, Astronomical
  Data Analysis Software and Systems XIV. p.~29

\bibitem[\protect\citeauthoryear{Taylor, Goss, Coleman, {van Leeuwen}  \&
  Wallace}{Taylor et~al.}{1996}]{Taylor1996}
Taylor A.~R.,  Goss W.~M.,  Coleman P.~H.,  {van Leeuwen} J.,   Wallace B.~J.,
  1996, \mn@doi [ApJS] {10.1086/192363}, 107, 239

\bibitem[\protect\citeauthoryear{Terzian}{Terzian}{1966}]{Terzian1966}
Terzian Y.,  1966, \mn@doi [ApJ] {10.1086/148646}, 144, 657

\bibitem[\protect\citeauthoryear{Terzian \& Dickey}{Terzian \&
  Dickey}{1973}]{Terzian1973}
Terzian Y.,  Dickey J.,  1973, AJ, 78, 875

\bibitem[\protect\citeauthoryear{Thomasson \& Davies}{Thomasson \&
  Davies}{1970}]{Thomasson1970}
Thomasson P.,  Davies J.~G.,  1970, MNRAS, \href
  {http://adsabs.harvard.edu/abs/1970MNRAS.150..359T} {150, 359}

\bibitem[\protect\citeauthoryear{Tylenda, Si{\'o}dmiak, G{\'o}rny, Corradi  \&
  Schwarz}{Tylenda et~al.}{2003}]{Tylenda2003}
Tylenda R.,  Si{\'o}dmiak N.,  G{\'o}rny S.~K.,  Corradi R. L.~M.,   Schwarz
  H.~E.,  2003, \mn@doi [A\&A] {10.1051/0004-6361:20030645}, 405, 627

\bibitem[\protect\citeauthoryear{Urquhart et~al.,}{Urquhart
  et~al.}{2009}]{Urquhart2009}
Urquhart J.~S.,  et~al., 2009, \mn@doi [A\&A] {10.1051/0004-6361/200912108},
  501, 539

\bibitem[\protect\citeauthoryear{{Van de Steene} \& Jacoby}{{Van de Steene} \&
  Jacoby}{2001}]{VandeSteene2001}
{Van de Steene} G.~C.,  Jacoby G.~H.,  2001, \mn@doi [A\&A]
  {10.1051/0004-6361:20010306}, 373, 536

\bibitem[\protect\citeauthoryear{Virtanen et~al.,}{Virtanen
  et~al.}{2020}]{virtanen2020}
Virtanen P.,  et~al., 2020, \mn@doi [Nature Methods, Vol. 17, pp. 261-272]
  {10.1038/s41592-019-0686-2}, 17, 261

\bibitem[\protect\citeauthoryear{Volk \& Kwok}{Volk \& Kwok}{1985}]{Volk1985}
Volk K.,  Kwok S.,  1985, A\&A, \href
  {http://adsabs.harvard.edu/abs/1985A\%26A...153...79V} {153, 79}

\bibitem[\protect\citeauthoryear{Wright, Griffith, Burke  \& Ekers}{Wright
  et~al.}{1994}]{Wright1994}
Wright A.~E.,  Griffith M.~R.,  Burke B.~F.,   Ekers R.~D.,  1994, \mn@doi
  [A\&AS] {10.1086/191939}, \href
  {http://adsabs.harvard.edu/abs/1994ApJS...91..111W} {91, 111}

\bibitem[\protect\citeauthoryear{Zijlstra, Pottasch  \& Bignell}{Zijlstra
  et~al.}{1989}]{Zijlstra1989a}
Zijlstra A.~A.,  Pottasch S.~R.,   Bignell C.,  1989, A\&AS, 79, 329

\bibitem[\protect\citeauthoryear{{van Hoof}}{{van Hoof}}{2000}]{Hoof2000a}
{van Hoof} P. A.~M.,  2000, MNRAS, 314, 99

\bibitem[\protect\citeauthoryear{{van de Steene} \& Pottasch}{{van de Steene}
  \& Pottasch}{1995}]{Steene1995}
{van de Steene} G.~C.,  Pottasch S.~R.,  1995, A\&A, \href
  {http://adsabs.harvard.edu/abs/1995A\%26A...299..238V} {299, 238}

\makeatother
\end{thebibliography}

%\bibliography{/Users/ibojicic/data/Biblio/papers_osx.bib}

%\begin{thebibliography}{99}
%\end{thebibliography}

%%%%%%%%%%%%%%%%%%%%%%%%%%%%%%%%%%%%%%%%%%%%%%%%%%

%%%%%%%%%%%%%%%%% APPENDICES %%%%%%%%%%%%%%%%%%%%%

% \appendix

% \input{appendixcontent} 

%%%%%%%%%%%%%%%%%%%%%%%%%%%%%%%%%%%%%%%%%%%%%%%%%%

% Don't change these lines
\bsp	% typesetting comment
\label{lastpage}
\end{document}